# Observation of Full-Parameter Jones Matrix in Bilayer Metasurface

Yanjun Bao[1,*], Fan Nan[1], Jiahao Yan[1], Xianguang Yang[1], Cheng-Wei Qiu[2,*] and Baojun Li[1,*]

[1]Institute of Nanophotonics, Jinan University, Guangzhou 511443, China

[2]Department of Electrical and Computer Engineering, National University of Singapore, Singapore 117583, Singapore

*Corresponding Authors: Y. Bao (yanjunbao@jnu.edu.cn), C.-W. Qiu (chengwei.qiu@nus.edu.sg), B. Li (baojunli@jnu.edu.cn)

**Abstract:** Metasurfaces, artificial 2D structures, have been widely used for the design of various functionalities in optics. Jones matrix, a 2×2 matrix with eight parameters, provides the most complete characterization of the metasurface structures in linear optics, and the number of free parameters (i.e., degrees of freedom, DOFs) in the Jones matrix determines the limit to what functionalities we can realize. Great efforts have been made to continuously expand the number of DOFs, and a maximal number of six has been achieved recently. However, the realization of "holy grail" goal with eight DOFs (full free parameters) has been proven as a great challenge so far. Here, we show that by cascading two layer metasurfaces and utilizing the gradient descent optimization algorithm, a spatially varying Jones matrix with eight DOFs is constructed and verified numerically and experimentally in optical frequencies. Such ultimate control unlocks new opportunities to design optical functionalities that are unattainable with previously known methodologies and may find wide potential applications in optical fields.

## Main

The design of optical structures with arbitrary functionalities has always been an ultimate dream for people in optics. However, the number of the degrees of freedom (DOFs) of light control by the implemented optical structure itself determines the limit to what functionalities that can be realized. In linear optics, the optical structures can be completely characterized by a spatially varying $2\times2$ Jones matrix[1] , which contains of eight parameters, therefore indicating a maximal number of eight DOFs (otherwise known as the free parameters that can be arbitrarily varied) for all linear structures in nature. Obviously, the more free parameters in the Jones matrix can be varied, the diverse functionalities we can achieve. The highest eight DOFs (full free parameters) represent the most general control in optics and are the foundation to realize the most complex optical functionalities. So, the question arises that how to expand the number of DOFs in Jones matrix, even to the ultimate eight DOFs?

Metasurfaces, which consist of a monolayer of planar structures, provide a suitable platform for spatially varying light control in subwavelength scales[2-4]. The metasurfaces have been extensively explored in the past ten years for diverse functionalities, such as anomalous refraction[4-8], hologram[9-16], metalens[17-22], vortex beam[23-26], etc. In fact, almost all these functionalities can be categorized into the different DOFs in the Jones matrix, which manifests the process of the continuous endeavors to expand the number of DOFs. For example, the $x$-polarized anomalous refraction with $y$-polarized incidence[4] can be attributed to the phase control of the $J_{12}$ component of Jones matrix, i.e., one DOF. The polarization-control dual holographic images[16] is enabled by the independent phase control of the two diagonal entries of Jones matrix, i.e., two DOFs. An application of four DOFs (the amplitude and phase terms of both $J_{12}$ and $J_{22}$ components in Jones matrix) is to generate arbitrary amplitude, phase, and polarization distributions[23]. Due to the mirror symmetry, the Jones matrix of planar structure is symmetric and thus has an upper limit number of six DOFs[27], which is constructed with metasurface recently[28, 29]. To break the mirror symmetry, multi-layer design is necessary. Recently, Yuan et al.[30] proposed a five-layer metallic structure to independently control the phases of the four components of the Jones matrix (circular polarization base) in microwave reign (i.e., four DOFs). Although great achievements have been made in expanding the number of DOFs, the "holy grail" goal with ultimate eight DOFs in Jones matrix has not yet been realized. The construction of such Jones matrix, especially operated in optical frequencies, is of great importance and meaningful to the field of optical design.

Here, we proposed and experimentally demonstrated an arbitrary spatially varying Jones matrix with eight DOFs in optical frequencies by using a bilayer metasurface to break the mirror symmetry of planar structures (Fig. 1a). The Jones matrix distributions of the two single layers in the bilayer structure are calculated based on gradient descent optimization algorithm, and the optimized results can agree well with any designed target distributions. The Jones matrix with eight DOFs provides unparalleled control of light. One example is that we can impose arbitrary and independent amplitude and phase control on any set of two polarizations (Fig. 1b). Most importantly, there are no restrictions on the input and output polarizations. In comparison, previously reported

Jones matrix with three DOFs can impart independent phase[31] (or amplitude[32]) control on orthogonal polarizations only, and the output polarizations must be the same as the input ones with flipped handedness (or mutually conjugate), i.e., the output two polarizations are orthogonal with each other, too. Following work with six DOFs[33] extends the above light control and can impart both independent amplitude and phase control, but exhibits the same restrictions on the input and output polarizations as that with three DOFs. The comparison in Fig. 1c highlights the unique and versatile control with eight DOFs. In addition, the bilayer design introduces another rotation DOF, which is utilized to realize polarization-rotation multiplexed holography reaching up to 16 independent functionalities.

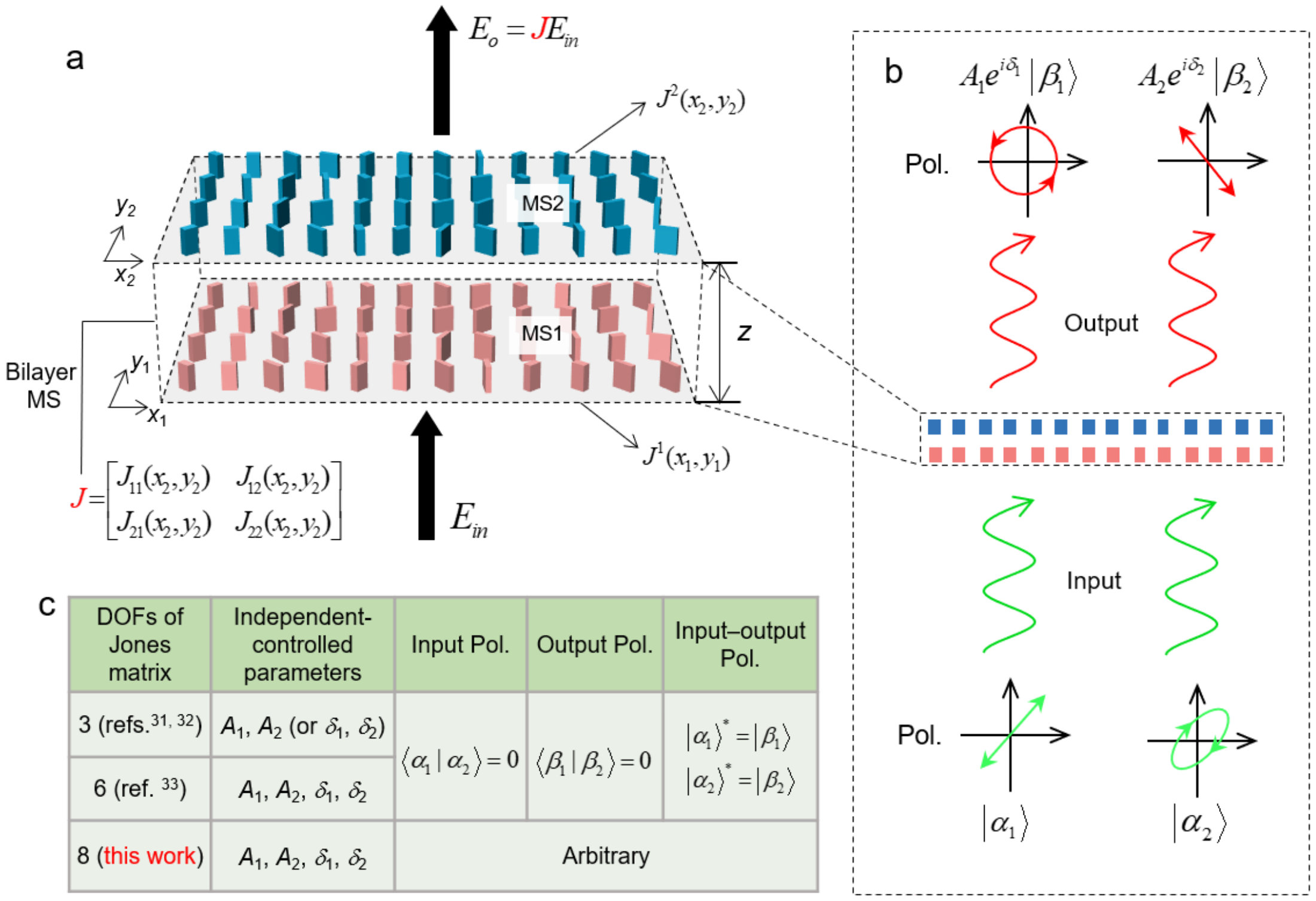

| DOFs of Jones matrix | Independent-controlled parameters | Input Pol. | Output Pol. | Input–output Pol. |
|---|---|---|---|---|
| 3 (refs.[31, 32]) | $A_1$, $A_2$ (or $\delta_1$, $\delta_2$) | $\langle\alpha_1\mid\alpha_2\rangle=0$ | $\langle\beta_1\mid\beta_2\rangle=0$ | $\vert\alpha_1\rangle^*=\vert\beta_1\rangle$ $\vert\alpha_2\rangle^*=\vert\beta_2\rangle$ |
| 6 (ref. [33]) | $A_1$, $A_2$, $\delta_1$, $\delta_2$ | | | |
| 8 (this work) | $A_1$, $A_2$, $\delta_1$, $\delta_2$ | Arbitrary | | |

**Fig. 1: Jones matrix with eight DOFs and an application example of advanced light control. a,** Schematic view of a bilayer metasurface with eight DOFs in the Jones matrix. The incident and output Jones vectors $E_{\mathrm{in}}$ and $E_{\mathrm{o}}$ are connected by 2×2 equivalent Jones matrix $J$. The two layer

metasurfaces are separated by a distance $z$. The Jones matrix of each single layer is symmetric and endowed with six DOFs. MS: metasurface. **b**, Schematic of the bilayer metasurface for independent amplitude and phase control of arbitrary set of two polarizations. The incident two arbitrary polarizations $|\alpha_1\rangle$ and $|\alpha_2\rangle$ can be transformed into arbitrary output polarizations $|\beta_1\rangle$ and $|\beta_2\rangle$ with independent complex-amplitude $A_1 e^{i\delta_1}$ and $A_2 e^{i\delta_2}$, respectively. There are no constraints imposed on the input and output polarizations. **c**, Comparison of the independent controlled parameters and polarization constraints between metasurfaces with different DOFs in Jones matrix.

Figure 1a illustrates the schematic view of our designed structure with eight DOFs in the Jones matrix, which consists of two layers of metasurfaces with a separation $z$ between them. Such bilayer metasurface design approaches have been previously used for other optical functionalities[34-39]. Here, the Jones matrix of each of the two layers $J^1(x_1, y_1)$ and $J^2(x_2, y_2)$ is symmetric due to mirror symmetry and assumed to have upper limit of six DOFs to provide enough design freedom. When light impinges from the bottom of the bilayer metasurface, it firstly passes through the first layer, then propagates over a distance $z$ in gap (homogenous environment) and finally passes through the second layer. The incident and output Jones vectors through the bilayer metasurface can be characterized by a spatially varying 2×2 Jones matrix $J$, which represents the optical properties of the whole optical system. In the following, we refer to it as equivalent Jones matrix (EQJM) in order to distinguish from that of the two single layers. The mn*th* ($m$, $n$=1, 2) component of the EQJM can be written as (see supplementary section 1)

$$J_{mn}(x_2,y_2)=\sum_{q=1,2} J^2_{mq}(x_2,y_2)\iint_{x_1,y_1} J^1_{qn}(x_1,y_1)\cdot f(x_2-x_1,y_2-y_1,z)dx_1dy_1 \quad (1)$$

where $J^1_{qn}$ and $J^2_{mq}$ are the *qn*th and *mq*th component of the Jones matrix of the first and second layer, respectively, $f(x_2-x_1,y_2-y_1,z)=\frac{1}{2\pi}\frac{\exp(ikr)}{r}\frac{z}{r}\left(\frac{1}{r}-i\frac{2\pi}{\lambda}\right)$ is the Rayleigh–Sommerfeld impulse response, $r=\sqrt{(x_1-x_2)^2+(y_1-y_2)^2+z^2}$, $i$ is the imaginary unit, $\lambda$ is the wavelength and *z* is the distance between the two layers. Note that the Jones matrix of single layer is symmetric and therefore $J^1_{12}(x_1,y_1)=J^1_{21}(x_1,y_1)$ and $J^2_{12}(x_2,y_2)=J^2_{21}(x_2,y_2)$.

We aim to find the Jones matrix values of the two single layers ($J^1_{ij}(x_1,y_1), J^2_{ij}(x_2,y_2)$, *ij*=11, 12 (21), 22) to design an arbitrary spatially varying EQJM with eight DOFs. The analytical expression in Equation (1) allows the convenient use of gradient descent optimization algorithm to obtain the optimized solutions (Fig. 2a and see supplementary section 2 for details). Besides the conventional absolute-mean-squared-error loss, we also add a boundary constraint loss $L_{bnd}$ to ensure that this algorithm converges to solutions that are inside the valid domains. The specific form of the boundary loss is determined based on the metasurface unit design, which will be introduced in the following. A detailed discussion of this boundary loss is provided in supplementary section 3.

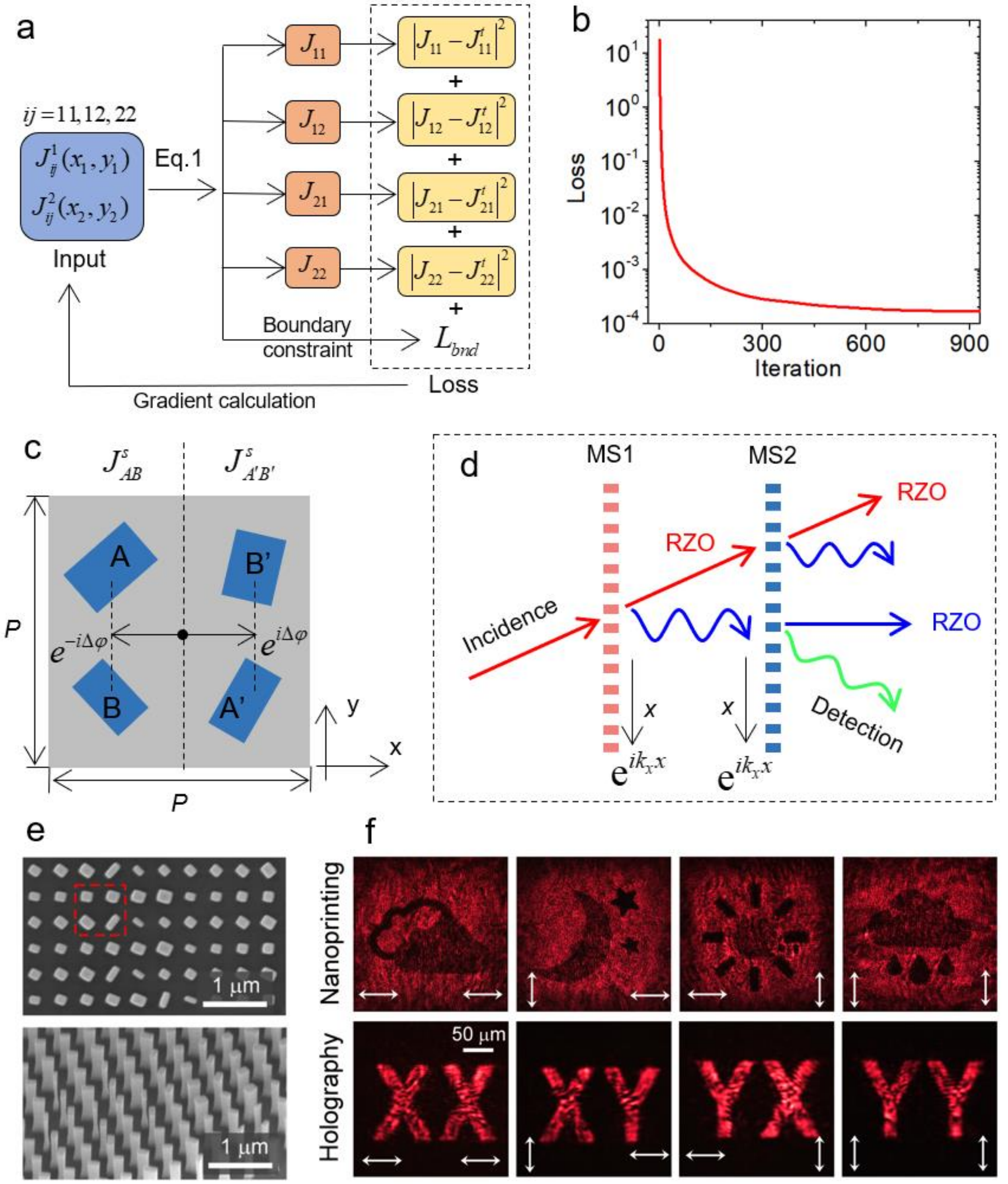

**Fig. 2: Design of Jones matrix with eight DOFs and measurement results. a**, Flow chart of the gradient descent optimization algorithm for design of Jones matrix with eight DOFs. The input variables are the Jones matrix components of the two single layer metasurface ( $J_{ij}^1(x_1, y_1), J_{ij}^2(x_2, y_2)$ , *ij*=11, 12 (21), 22). The defined loss includes two parts: the mean of the absolute squared differences of EQJM components between the prediction ( $J_{ij}$ ) and target ( $J_{ij}^t$ ), and a boundary constraint loss $L_{bnd}$ . **c**, Loss value as a function of the iteration number in the gradient descent optimization algorithm. **c**, Unit cell of the single layer metasurface to construct Jones matrix with six DOFs. Each unit contains of four dielectric nanopillars, and the two sets of nanopillars *AB* and *A*'*B*' can individually and independently construct Jones matrix with six DOFs. Under oblique incidence or oblique scattering, detour phases of $e^{-i\Delta\varphi}$ and $e^{i\Delta\varphi}$ are imposed on the two sets of nanopillars *AB* and *A*'*B*'. To compensate the detour phases, the Jones matrixes of the two set of nanopillars are chosen as $J_{AB}^{s}=e^{i\Delta\varphi}J$ and $J_{A'B'}^{s}=e^{-i\Delta\varphi}J$ to constructively generate the designed Jones matrix *J* for the unit cell. The period of the squared unit pixel is *P*=800 nm. **d**, Schematic of light propagation through the bilayer metasurface with the oblique incidence-oblique detection measurement strategy. For each layer, the scattering light (indicated by wavy line arrows) is deflected by applying a gradient phase distribution $e^{ik_x x}$ on it and the residual zero order maintains its propagation direction as the incident one. With such measurement, the influence of the residual zero orders can be totally eliminated in the detection. RZO: residual zero order. **e**, Scanning electron microscopy (SEM) images of the fabricated metasurface (partial view). The dashed red square indicates one unit pixel with individually designing the two sets of nanopillars. **f,** Experimental results of the measured nanoprinting and holographic images. The incident and analyzed polarizations are indicated at lower left and lower right of each panel. All panels share the same scale bar in the lower left one.

To verify our approach, a target EQJM is designed with its amplitude and phase distributions chosen to present four nanoprintings (trinary intensity images of weather symbol) and four holographic images (letter strings “XX”, “XY”, “YX” and “YY”) encoded in its four components (The details can be found in supplementary section 4).

Figure 2b plots the loss value obtained from the algorithm as a function of the iteration number, which decays rapidly and reaches convergence after nearly 600 iterations. The optimized results of the EQJM including both the amplitude and phase distributions show good agreements with the targets (supplementary figure 2). Only a slight deviation of the magnitude occurs near the boundaries, where the sharp truncation of the magnitude contains high angular spectrum frequencies that extend beyond our design.

The two single layer metasurface with six DOFs of Jones matrix can be constructed by multi-element unit design[28, 29]. In this case, the Jones matrix is decomposed as the summation of the ones of the individual elements. We consider the element with rectangle dielectric nano pillars of silicon on glass, which have a higher refractive index than the surrounding environment air. The Jones matrix of the single layer metasurface with two-element (nanopillars *A* and *B*, Fig. 2c) unit is given by

$$J_{AB}^{s}=\begin{bmatrix} J_{11}^{s} & J_{12}^{s} \\ \sim & J_{22}^{s} \end{bmatrix}=R(-\theta_A)\begin{bmatrix} e^{i\varphi_1} & 0 \\ 0 & e^{i\varphi_2} \end{bmatrix}R(\theta_A)+R(-\theta_B)\begin{bmatrix} e^{i\varphi_3} & 0 \\ 0 & e^{i\varphi_4} \end{bmatrix}R(\theta_B) \tag{2}$$

where $\varphi_1$ ($\varphi_3$) and $\varphi_2$ ($\varphi_4$) are the phase shifts imposed on the light linearly polarized along the fast and slow axes of nanopillar *A* (*B*), $\theta_{A,B}$ are the rotational angles of the two nanopillars and $R(\theta_{A,B})$ is the $2\times2$ rotation matrix. Clearly, not all symmetric Jones matrix can be decomposed as that in Eq. (2). A sufficient precondition of Eq. (2) is $|J_{11}^{s}|+|J_{12}^{s}|\le 2$ and $|J_{12}^{s}|+|J_{22}^{s}|\le 2$. The details of the derivations are provided in supplementary section 3. Therefore, a boundary constraint loss is added in the gradient descent algorithm to ensure that the input Jones matrix values always fall inside the above domains.

The phase shift values $\varphi_1$ ($\varphi_3$) and $\varphi_2$ ($\varphi_4$) are associated with the transverse dimensions (length and width in *xy* plane) of the nanopillar. Full wave finite-difference time-domain (FDTD) simulations are performed, and a library of the transmission magnitudes and phase shifts dependent on the transverse dimensions of the nanopillar with incident *x*- and *y*- polarizations is build. With such databases, any $\varphi_1$ ($\varphi_3$) and $\varphi_2$ ($\varphi_4$) combinations ranging from 0 to $2\pi$ can be achieved by properly selecting the transverse dimensions of the nanopillars (see details in supplementary section 4).

The two nanopillars *A* and *B* are arranged vertically along *y* axis, and to create a square pixel, a simple way is to duplicate the set of nanopillars *AB* to *A'B'* and distribute them uniformly within the pixel (Fig. 2c). This treatment is appropriate under normal incidence and normal scattering. For oblique incidence or oblique scattering (along *x* direction), the introduced detour phases of the two sets of nanopillars can cause a major reduction of efficiency and deteriorate the optical performance (see discussion in supplementary section 7). It is necessary to individually design the optical responses of the two set of nanopillars *AB* and *A'B'* under this scenario.

When the incident light passes through the first layer metasurface, besides the transmitted scattering field, the unwanted residual zero-order light also imposes on the second layer. The two beams then pass through the second layer metasurface and respectively generate both the scattered field and residual zero order. Here the field of interest only is the scattering from the second layer illuminated by the scattering from the first layer. The residual zero-order light may strongly affect the measurement when its magnitude is comparable to that of the designed images.

We apply an oblique incidence-oblique detection measurement strategy, as shown in Fig. 2d. Here, the light is obliquely incident on the first layer, and the transmitted scattering is bended towards the normal direction by applying a gradient phase distribution $e^{ik_x x}$ ($k_x = 0.3k$, $k$ is the wavevector of the light in air) on the first layer. The same gradient phase distribution is also added on the second layer, which diffracts the incident normal scattering to an oblique direction. The influence of the residual zero orders can be totally eliminated if only the obliquely diffracted scattering is collected for imaging. In each pixel design, the Jones matrix of the nanopillars *AB* and *A'B'* are set as $e^{i\Delta\varphi}J$ and $e^{-i\Delta\varphi}J$ to compensate the detour phases, where $\Delta\varphi$ is the detour phase arisen from the oblique incidence or oblique scattering and $J$ is the designed Jones matrix value for the pixel. The advantages of the above measurement strategy are demonstrated (see supplementary fig. 5) by performing the full-wave FDTD simulations of the realistic bilayer metasurface under different optical measurement setups (normal incidence-normal detection, oblique incidence-normal detection, and oblique incidence-oblique detection). The details of the simulation processes are provided in supplementary section 6. It is noted that the image quality of the holographic image maintains high fidelity under all measurement setups. This is mainly because that the holographic image is designed to be highly focused in far field with its magnitude much larger than the diffracted background noise of the residual zero orders. Therefore, if one is only interested in the highly focused field manipulation away from the metasurface plane[28], the convenient normal incidence-normal detection strategy can be used. Additionally, a comparison between the consideration of the detour phase in the unit pixel design and without is shown in supplementary fig. 7. We also investigate the optical performance on the alignment of the two layers, which reveals that the images can be recognized with translational movement shift less than about 5 μm (supplementary fig. 8).

To set up the bilayer metasurface, we utilized two metasurfaces on separated substrates and cascaded them front-to-front to maintain a homogeneous environment (air) between them. The two metasurface samples are fabricated on 600-nm-height crystal silicon layer that is transferred on glass substrate. The patterns are then defined by electron beam lithography (EBL) and reactive ion etching (RIE) process. The details of the fabrication procedure are outlined in the Methods. Figure 2e shows the SEM images of the metasurfaces, where the different designs of nanopillar sets of *AB* and *A*'*B*' within the pixel are observed. The oblique view of the metasurface shows the smooth sidewall profiles of these nanopillars. To measure the optical images along oblique direction, we carry out a spatial filtering process in the Fourier plane. The Fourier plane of the objective lens (usually lies inside its barrel) is taken outside by adopting two lenses, which is then readily available for the filtering process. The details of the optical measurement are shown in Methods and supplementary section 9.

The experimental results of the measured optical images with different combinations of incident and analyzed polarizations are shown in Fig. 2f, which have good agreements with the designed targets. Although the measurement of the nanoprintings can be easily influenced by optical setups, the trinary intensity distributions are clearly seen for all four nanoprintings, demonstrating the accurate amplitude control of the EQJM. Due to the phase variations (arising from oblique incidence-oblique detection), the measured nanoprinting images exhibit fringe patterns, which are also verified in the simulations (supplementary fig. 5).

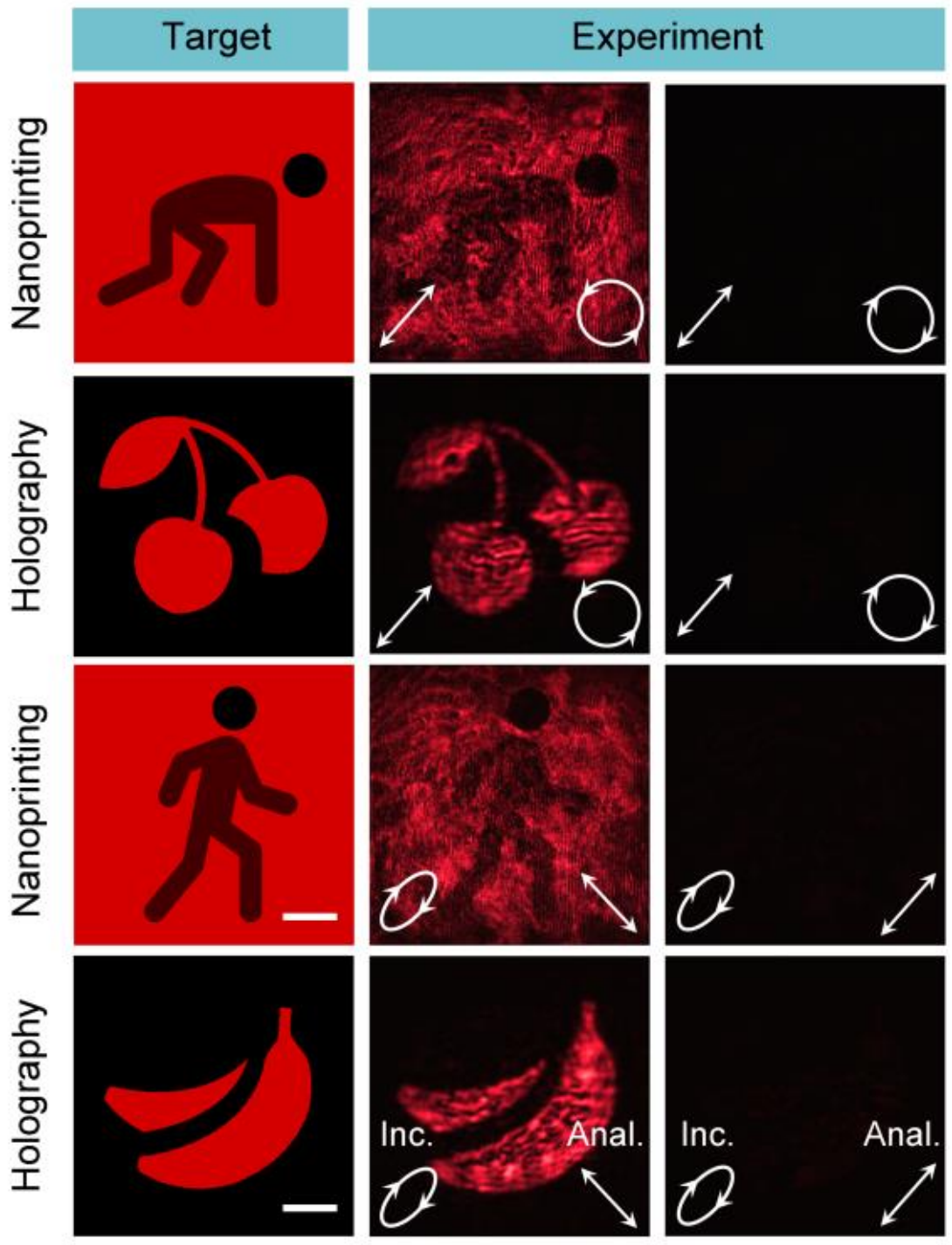

**Fig. 3: Experiment demonstration of the independent amplitude and phase control of arbitrary set of two polarizations based on Jones matrix with eight DOFs.** The figure shows the target and measured nanoprintings and holographic images. The incident and analyzed polarizations are indicated in lower left and lower right corners of each panel. Scale bar: 50 μm.

Next, we aim to demonstrate the proposed light control enabled by the eight DOFs that is mentioned in Fig. 1b. Let the two incident arbitrary polarizations be given by Jones vectors $|\alpha_1\rangle = \begin{bmatrix} \gamma_1 \\ \gamma_2 \end{bmatrix}$ and $|\alpha_2\rangle = \begin{bmatrix} \gamma_3 \\ \gamma_4 \end{bmatrix}$. The metasurface transforms the two input polarizations to output polarizations $|\beta_1\rangle = \begin{bmatrix} \chi_1 \\ \chi_2 \end{bmatrix}$ and $|\beta_2\rangle = \begin{bmatrix} \chi_3 \\ \chi_4 \end{bmatrix}$ with independent complex-amplitude control of $A_1 e^{i\delta_1}$ and $A_2 e^{i\delta_2}$, respectively (Fig. 1b). Note that we do not impose any constraints on the input and output polarizations. The Jones matrix of the metasurface should simultaneously satisfy

$$J|\alpha_1\rangle = A_1 e^{i\delta_1}|\beta_1\rangle \tag{3}$$

and

$$J|\alpha_2\rangle = A_2 e^{i\delta_2}|\beta_2\rangle \tag{4}$$

The Jones matrix can be directly extracted as

$$J = \begin{bmatrix} A_1 e^{i\delta_1}\chi_1 & A_2 e^{i\delta_2}\chi_3 \\ A_1 e^{i\delta_1}\chi_2 & A_2 e^{i\delta_2}\chi_4 \end{bmatrix} \begin{bmatrix} \gamma_1 & \gamma_3 \\ \gamma_2 & \gamma_4 \end{bmatrix}^{-1} \tag{5}$$

Obviously, we can always use the bilayer metasurface to construct a Jones matrix that satisfies Eq. (5) only if the input two polarizations are not exactly the same. It is worth mentioning that if only one of the two conditions (Eq.3 and 4) is satisfied, i.e., converting arbitrary polarization into another polarization state with independent amplitude and phase control, the Jones matrix requires a minimal number of four DOFs[40]. For demonstration, we choose nonorthogonal linear polarization $(\gamma_1 = \frac{\sqrt{2}}{2}, \gamma_2 = \frac{\sqrt{2}}{2})$ and elliptical polarization $(\gamma_3 = \frac{\sqrt{2}}{2}, \gamma_4 = \frac{\sqrt{2}}{2} e^{i\pi/3})$ as the input ones, which are transformed into another nonorthogonal circular polarization $(\chi_1 = \frac{\sqrt{2}}{2}, \chi_2 = -\frac{\sqrt{2}}{2} i)$ and linear polarization $(\chi_3 = \frac{\sqrt{2}}{2}, \chi_4 = -\frac{\sqrt{2}}{2})$, respectively. The amplitude and phase control are demonstrated by designing two nanoprintings and two holographic images (first column in Fig. 3). When the incident and analyzed polarizations are set as the designed ones, we can observe clear nanoprintings and holographic images in the measurement (second column in Fig. 3), agreeing well with the targets. The output polarizations are verified from the observed almost dark images when switching the analyzed polarizations to the orthogonal ones (third column in Fig. 3).

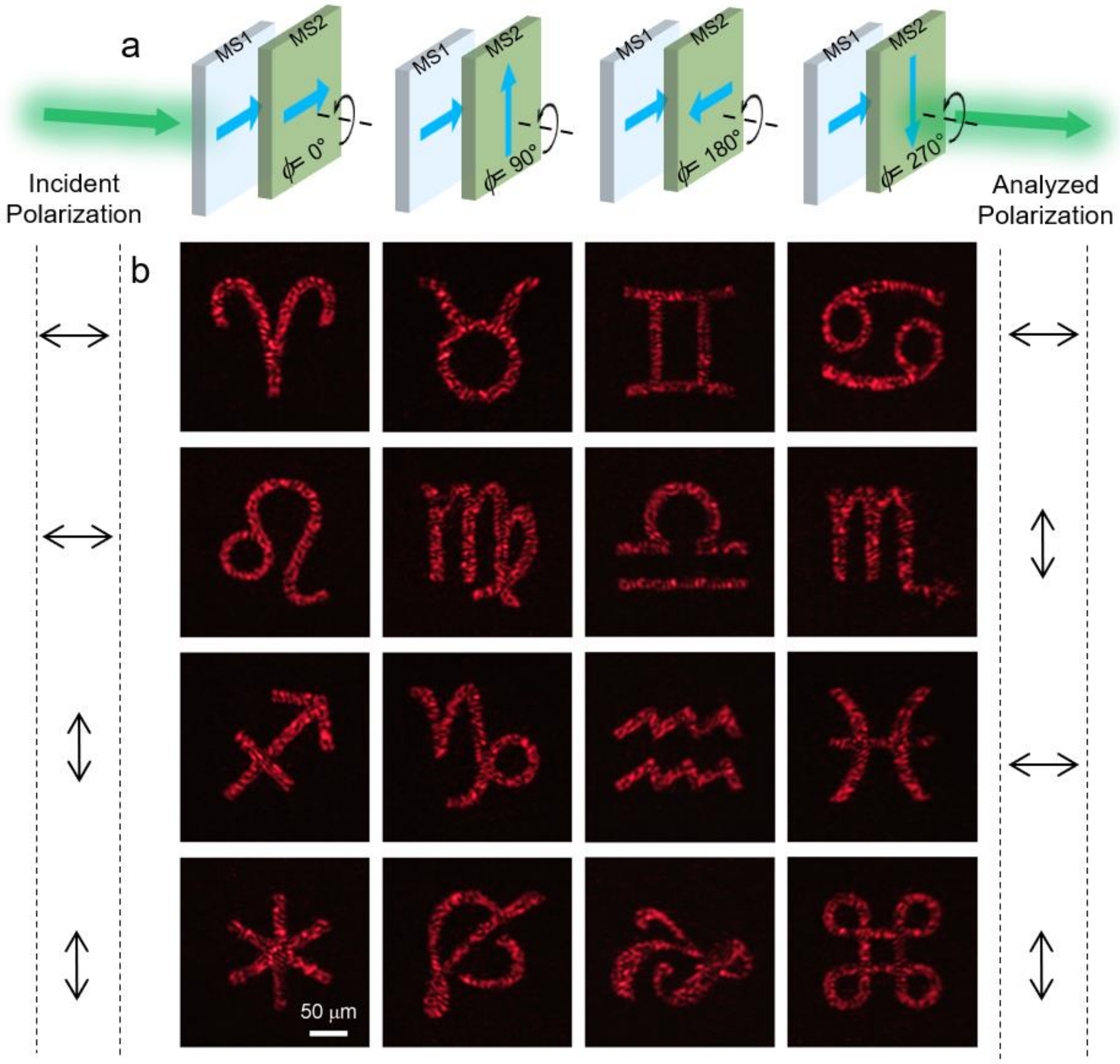

**Fig. 4: Bilayer metasurface for polarization-rotation multiplexed multifunctional holography.** **a**, Schematic of the two layer metasurfaces with different rotation angle $\phi$= 0°, 90°, 180° and 270°. The arrows on the metasurfaces show the relative rotation angles between the two layers. **b**, Measured holographic images under the four different combinations of the incident and analyzed polarizations with rotation angles of $\phi$= 0°(first row), 90°(second row), 180°(third row) and 270°(fourth row). The holographic images are designed at 1500 μm above the second layer in the glass substrate. The incident and analyzed polarizations are indicated at left side and right side, respectively. All panels share the same scale bar in the lower left one.

The bilayer metasurface design introduces another DOF, the rotation angle between the two layers (fig. 4a). It can be imagined that when the two layer metasurfaces rotate with respect to each other, the whole structure will exhibit different optical responses, which are that we want to control. The proposed bilayer metasurface is a good platform for multifunctional control as it provides enough DOFs (12 $N^2$, $N$ is the pixel number along one direction) for design. We consider four cases with rotational angles $\phi$= 0°, 90°, 180° and 270°, and in each case, four independent functionalities are designed for each of the four different combinations of the incident and analyzed polarizations, i.e., a total of 16 polarization-rotation multiplexed functionalities. Here, we only consider the holographic functionality, as it is robust to the measurement conditions (see supplementary fig. 5). The loss is defined as the summation of the 16 squared differences between the magnitudes the predicted holographic images and the targets, plus the boundary loss (see details in supplementary section 10). Then the gradient descent optimization algorithm is used to retrieve the Jones matrix values of the two single layers. To avoid the overlapping of the holographic image and the possible residual zero order, we applied the oblique incidence-normal detection measurement strategy. Figure 4b display our measured 16 holographic images under different combinations of the four rotation angles, two incident polarizations and two analyzed polarizations. All the measured images have high fidelity and have good agreement with the calculated optimized results (supplementary fig. 10). More importantly, the measured results almost have no cross-talk between any two holographic images, demonstrating the full independent multifunctionalities.

Although the whole device in the demonstration is not that compact (the gap distance between the two layers is set as 150 μm for the measurement convenience), our design strategy is general, and can be extended to gap distance towards several wavelengths (e.g., 5 μm), as shown in supplementary section 11. Further efforts can be made to integrate the two layer metasurfaces on the two sides of one substrate[38, 39]. It is noteworthy that although the demonstration was designed at 808 nm, which is aimed to avoid the large optical loss of silicon, our approach applies to short wavelengths in the visible range and can maintain performance with other lossless materials, such as $TiO_2$.

In conclude, we have cascaded two single layer metasurfaces with six DOFs in the Jones matrix to construct a spatially varying Jones matrix with full parameters of eight DOFs, the maximal number allowed in nature. This represents a significant advance in the state of the art for light control. Enabled by the eight DOFs of light control, we have demonstrated novel functionalities with independent amplitude and phase control of arbitrary set of two polarizations, without any constraints on the incident and output polarizations. In addition, we have investigated the DOF of rotation in the bilayer metasurface and demonstrated polarization-rotation multiplexed holography reaching up to 16 independent functionalities. We believe that our proposed design strategy of eight DOFs in Jones matrix offers a generalize method towards arbitrary control of light and may find novel applications that are not attainable with conventional methods.

## Methods

**Sample fabrication.** A commercial SOI wafer with a 1200 nm-thickness of device layer is firstly transferred on glass substrate by adhesive wafer bonding and deep reactive ion etching (DRIE). The thickness of the device layer is further reduced to 600 nm using inductively coupled plasma (ICP). To fabricate the metasurface pattern, a 300 nm-thickness hydrogen silsesquioxane (HSQ) layer is first spin-coated at 4000 rpm on the substrate and baked on a hot plate for 5 min at 90 °C. Then a 30 nm thickness aluminum layer (thermal evaporation) is deposited to serve as the charge dissipation layer. Next, the pattern is exposed using electron beam lithography (EBL). After exposure, the aluminum layer is removed by 5% phosphoric acid, and the resist is developed with tetramethy- lammonium hydroxide. Finally, the sample is etched using ICP. An important note is that the two metasurfaces are cascaded front-to-front, and therefore the EBL pattern of the second metasurface should be flipped horizontally.

**Optical setup and measurement.** A schematic of the optical setup for the experimental measurement is shown in supplementary fig. 9. A tunable laser source is used to generate the light beam with a wavelength of 808 nm. The laser source is collimated with uniform intensity in the center and obliquely incident on the bilayer metasurface. The different polarization is generated by a polarizer and a quarter waveplates (QWP) in front of two metasurface samples, which are separately mounted on 3D translational stages. The light scattered by the bilayer metasurface is collected by a 20×/0.50 objective and isolated with another pair of QWP and polarizer. To measure the optical images along oblique direction, we carry out a spatial filtering process in the Fourier plane. The Fourier plane of the objective lens (usually lies inside its barrel) is taken outside by adopting Lenses 1 and 2, which is then readily available for the filtering process. A continuously variable iris is placed at the Fourier plane to serve as the filter. The position and the diameter of the iris are determined according to designed parameters (See details in supplementary section 9). The final image is focused by Lens 3 and formed on the CMOS camera.

**Acknowledgements**

This research was supported by the National Natural Science Foundation of China (92150107, 62075246) and Guangdong Natural Science Funds (2022B1515020019).

**Author contributions**

Y. B. conceived the idea, conducted the numerical simulations, fabricated the sample, performed the measurement and wrote the manuscript. X. Y., F. N., J. Y. and C-W. Q. joined the discussions and gave useful suggestions. Y. B., C-W. Q. and B. L. supervised the project.

**Competing interests**

The authors declare no competing interests.

# Supplementary Information

## Table of Contents

## S1. Derivation of Eq. (1)

We consider an optical system composed of two layer metasurfaces that are separated by a distance $z$ (Figure S1). The distributions of the Jones matrix values are denoted as $J^1 = \begin{bmatrix} J_{11}^1(x_1, y_1) & J_{12}^1(x_1, y_1) \\ J_{21}^1(x_1, y_1) & J_{22}^1(x_1, y_1) \end{bmatrix}$ and $J^2 = \begin{bmatrix} J_{11}^2(x_2, y_2) & J_{12}^2(x_2, y_2) \\ J_{21}^2(x_2, y_2) & J_{22}^2(x_2, y_2) \end{bmatrix}$ for metasurfaces 1 and 2, respectively, where ($x_1$, $y_1$) and ($x_2$, $y_2$) are the $xy$ coordinates. As the single metasurface layer has mirror symmetry with respect to the transverse plane, the off-diagonal elements of the Jones matrix should be identical, i.e, $J_{12}^1(x_1, y_1) = J_{21}^1(x_1, y_1)$ and $J_{12}^2(x_2, y_2) = J_{21}^2(x_2, y_2)$.

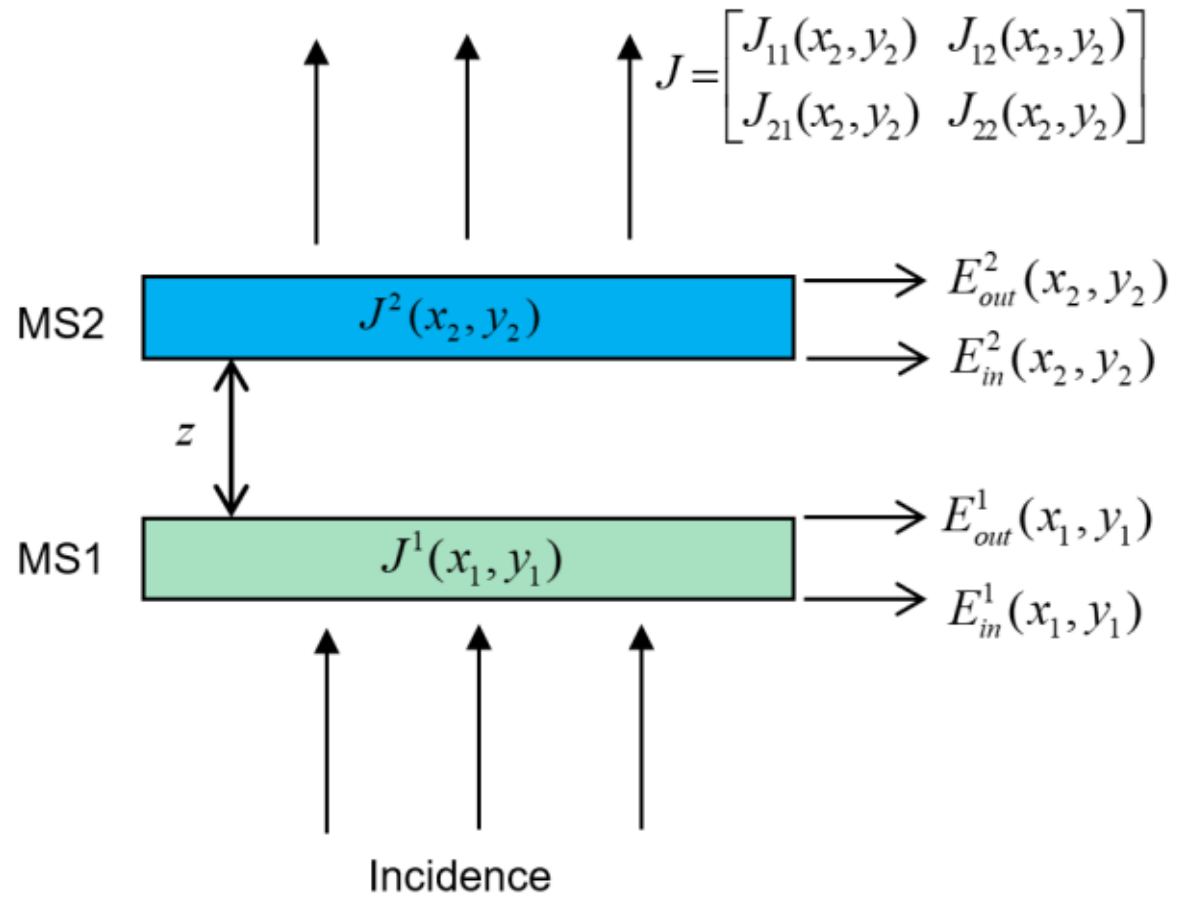

Figure S1. Schematic view of a two layer metasurface system separated by a vertical distance $z$. The Jones matrixes of the two single layers are denoted as $J^1(x_1, y_1)$ and $J^2(x_2, y_2)$, respectively. The light is incident from the bottom of metasurface 1, with the electric fields at the incident and output planes of metasurface 1 denoted by $E_{in}^1(x_1, y_1)$ and $E_{out}^1(x_1, y_1)$, the electric fields at the incident and output planes of metasurface 2 denoted by $E_{in}^2(x_2, y_2)$ and $E_{out}^2(x_2, y_2)$, respectively. $J$ represents the equivalent Jones matrix of the whole structure. MS: metasurface.

We then derive the equivalent Jones matrix expression $J(x_2, y_2)$ of the whole two-layer optical system. Considering the light is firstly incident on the metasurface 1 with polarization along $x$ direction $E_{in}^1 = [1\ \ 0]^T$, the electric field distributions passing through metasurface 1 is

$$E_{out}^1(x_1, y_1) = \begin{bmatrix} J_{11}^1(x_1, y_1) & J_{12}^1(x_1, y_1) \\ J_{21}^1(x_1, y_1) & J_{22}^1(x_1, y_1) \end{bmatrix} \begin{bmatrix} 1 \\ 0 \end{bmatrix} = \begin{bmatrix} J_{11}^1(x_1, y_1) \\ J_{21}^1(x_1, y_1) \end{bmatrix} \tag{S1.1}$$

After propagating a distance $z$, the electric field distributions $E_{in}^2(x_2, y_2)$ at the bottom surface of metasurface 2 is

$$E_{in}^2(x_2, y_2) = \begin{bmatrix} \iint\limits_{x_1, y_1} J_{11}^1(x_1, y_1) \cdot f(x_2 - x_1, y_2 - y_1, z) dx_1 dy_1 \\ \iint\limits_{x_1, y_1} J_{21}^1(x_1, y_1) \cdot f(x_2 - x_1, y_2 - y_1, z) dx_1 dy_1 \end{bmatrix} \tag{S1.2}$$

where we have used the Rayleigh–Sommerfeld diffraction formula to calculate the light propagation, $f(x_2 - x_1, y_2 - y_1, z) = \frac{1}{2\pi} \frac{\exp(ikr)}{r} \frac{z}{r} \left( \frac{1}{r} - i \frac{2\pi}{\lambda} \right)$ is the Rayleigh–Sommerfeld impulse response, $r = \sqrt{(x_1 - x_2)^2 + (y_1 - y_2)^2 + z^2}$, $i$ is the imaginary unit, $\lambda$ is the wavelength and $z$ is the distance between the two layers.

The electric field passing through metasurface 2 at the output plane is

$$\begin{aligned} E_{out}^2(x_2, y_2) &= J^2(x_2, y_2) \cdot E_{in}^2(x_2, y_2) \\ &= \begin{bmatrix} \sum\limits_{q=1,2} J_{1q}^2(x_2, y_2) \iint\limits_{x_1, y_1} J_{q1}^1(x_1, y_1) \cdot f(x_2 - x_1, y_2 - y_1, z) dx_1 dy_1 \\ \sum\limits_{q=1,2} J_{2q}^2(x_2, y_2) \iint\limits_{x_1, y_1} J_{q1}^1(x_1, y_1) \cdot f(x_2 - x_1, y_2 - y_1, z) dx_1 dy_1 \end{bmatrix} \end{aligned} \tag{S1.3}$$

Similarly, we can obtain the electric fields at the output plane of metasurface 2 with $y$-polarized incidence $E_{in}^1 = [0\ 1]^T$ as

$$E_{out}^2(x_2, y_2) = \begin{bmatrix} \sum\limits_{q=1,2} J_{1q}^2(x_2, y_2) \iint\limits_{x_1, y_1} J_{q2}^1(x_1, y_1) \cdot f(x_2 - x_1, y_2 - y_1, z) dx_1 dy_1 \\ \sum\limits_{q=1,2} J_{2q}^2(x_2, y_2) \iint\limits_{x_1, y_1} J_{q2}^1(x_1, y_1) \cdot f(x_2 - x_1, y_2 - y_1, z) dx_1 dy_1 \end{bmatrix} \tag{S1.4}$$

Based on Eqs. S1.3 and 1.4, we can obtain the equivalent Jones matrix of the two layer system with its *mn* (*m*, *n*=1, 2) component as

$$J_{mn}(x_2,y_2)=\sum_{q=1,2} J_{mq}^{2}(x_2,y_2)\iint_{x_1,y_1} J_{qn}^{1}(x_1,y_1)\cdot f(x_2-x_1,y_2-y_1,z)dx_1dy_1 \qquad \text{(S1.5)}$$

This is the derivation of Eq. 1 in the main text.

Alternatively, one can use the angular spectrum method to calculate the field propagation between the two layers. For this case, Equation S1.5 becomes

$$J_{mn}(x_2,y_2)=\sum_{q=1,2} J_{mq}^{2}(x_2,y_2)\iint_{u,v} A_{qn}^{1}(u,v,z)\cdot \exp(i2\pi(ux_2+vy_2))dudv \qquad \text{(S1.6)}$$

where $A_{qn}^{1}(u,v,z)=A_{qn}^{1}(u,v,0)G(u,v,z)$ , $G(u,v,z)=\exp(i2\pi z(\lambda^{-2}-u^2-v^2)^{1/2})$

and $A_{qn}^{1}(u,v,0)=\iint_{x_1,y_1} J_{qn}^{1}(x_1,y_1)\cdot \exp(-i2\pi(ux_1+vy_1))dx_1dy_1$ .

The two methods are equivalent. However, for numerical calculation, errors are introduced due to the discretization of the functions. The two methods are applicable to in different scenes. For example, for the same sampling intervals, the Rayleigh–Sommerfeld diffraction is suitable for long distance propagation[1] while the angular spectrum method is suitable for near field regions[2]. For our structures, the metasurface has a length size of hundred micrometers and sampled interval of 0.8 μm. If we want to calculate the light propagation between the two layers (with a gap distance of 150 micrometers), the angular spectrum method is more suitable. For the calculation of the holography which is designed at a distance of few thousand of micrometers above the metasurface 2, the Rayleigh–Sommerfeld diffraction method is more suitable. As the expression of the Rayleigh–Sommerfeld diffraction is more concise, we use it in the main text and the following derivations.

## S2. The detail of the gradient descent optimization

In this section, we provide the detail of the gradient descent optimization method to obtain the Jones matrix distributions of the two layer metasurfaces $J^1(x_1, y_1)$ and $J^2(x_2, y_2)$ to design a target equivalent Jones matrix distribution $J^t(x_2, y_2)$. For numerical calculations, all the planes are sampled to $N \times N$ equidistant grids with sampling intervals of $\Delta x = \Delta y = P$. We define a loss of

$$L_{mase} = \frac{1}{4N^2} \sum_{m=1}^{2} \sum_{n=1}^{2} \sum_{i=1}^{N} \sum_{j=1}^{N} \left| J_{mn}(x_{2i}, y_{2j}) - J_{mn}^t(x_{2i}, y_{2j}) \right|^2 \quad \text{(S2.1)}$$

which is the mean of the absolute squared error between the target $J_{mn}^t$ and the calculated Jones matrixes $J_{mn}$ of all the four components. Here, $J_{mn}$ is a function of the Jones matrix components of the two single layers (see Eq. S1.5 or Eq. S1.6). The core step of the gradient descent optimization method is to calculate the gradient of the defined loss with respect to the input variables, i.e., the Jones matrix components of the two single layers. Note that all the Jones matrix components are complex values and therefore have two independent variables. Take the *11*th component of the Jones matrix of the first layer metasurface for example, it is written as $J_{11}^1(x_{1p}, y_{1q}) = u_{11}^1(x_{1p}, y_{1q}) + i v_{11}^1(x_{1p}, y_{1q})$ and the gradient of $L_{mase}$ with respect to $u_{11}^1(x_{1i}, y_{1j})$ and $v_{11}^1(x_{1i}, y_{1j})$ can be calculated as

$$\frac{\partial L_{mase}}{\partial u_{11}^1(x_{1p}, y_{1q})} = 2\,\mathrm{Re}\left( \frac{\partial L_{mase}}{\partial J_{11}^1(x_{1p}, y_{1q})} \frac{\partial J_{11}^1(x_{1p}, y_{1q})}{\partial u_{11}^1(x_{1p}, y_{1q})} \right) = 2\,\mathrm{Re}\left( \frac{\partial L_{mase}}{\partial J_{11}^1(x_{1p}, y_{1q})} \right)$$

$$= 2\,\mathrm{Re}\left( \sum_{m=1}^{2} \sum_{n=1}^{2} \sum_{i=1}^{N} \sum_{j=1}^{N} \frac{1}{4N^2} \left( J_{mn}(x_{2i}, y_{2j}) - J_{mn}^t(x_{2i}, y_{2j}) \right)^* \frac{\partial J_{mn}(x_{2i}, y_{2j})}{\partial J_{11}^1(x_{1p}, y_{1q})} \right)$$

$$= 2\,\mathrm{Re}\left( \sum_{m=1}^{2}\sum_{i=1}^{N}\sum_{j=1}^{N} \frac{1}{4N^2} \left( J_{m1}(x_{2i}, y_{2j}) - J_{m1}^{t}(x_{2i}, y_{2j}) \right)^{*} \frac{\partial J_{m1}(x_{2i}, y_{2j})}{\partial J_{11}^{1}(x_{1p}, y_{1q})} \right)$$

$$= \frac{1}{2N^2}\mathrm{Re}\left( \sum_{m=1}^{2}\sum_{i=1}^{N}\sum_{j=1}^{N} \left( J_{m1}(x_{2i}, y_{2j}) - J_{m1}^{t}(x_{2i}, y_{2j}) \right)^{*} J_{m1}^{2}(x_{2i}, y_{2j}) f(x_{2i} - x_{1p}, y_{2j} - y_{1q}, z)\Delta x \Delta y \right)$$

$$\frac{\partial L_{mase}}{\partial v_{11}^{1}(x_{1p}, y_{1q})}$$

$$= \frac{1}{2N^2}\mathrm{Re}\left( i\sum_{m=1}^{2}\sum_{i=1}^{N}\sum_{j=1}^{N} \left( J_{m1}(x_{2i}, y_{2j}) - J_{m1}^{t}(x_{2i}, y_{2j}) \right)^{*} J_{m1}^{2}(x_{2i}, y_{2j}) f(x_{2i} - x_{1p}, y_{2j} - y_{1q}, z)\Delta x \Delta y \right)$$

where $p$, $q$ are the coordinate indexes of first layer and $i$, $j$ are the coordinate indexes of the second layer.

As $J_{12}^{1}(x_1, y_1) = J_{21}^{1}(x_1, y_1)$, the gradient of $F$ with respect to $J_{12}^{1}$ (or $J_{21}^{1}$) should be the sum of the gradients with respect to $J_{12}^{1}$ and $J_{21}^{1}$, that is

$$\frac{\partial L_{mase}}{\partial u_{12}^{1}(x_{1p}, y_{1q})}$$

$$= 2\,\mathrm{Re}\left( \frac{\partial F}{\partial J_{12}^{1}(x_{1p}, y_{1q})} \frac{\partial J_{12}^{1}(x_{1p}, y_{1q})}{\partial u_{12}^{1}(x_{1p}, y_{1q})} \right) + 2\,\mathrm{Re}\left( \frac{\partial F}{\partial J_{21}^{1}(x_{1p}, y_{1q})} \frac{\partial J_{21}^{1}(x_{1p}, y_{1q})}{\partial u_{21}^{1}(x_{1p}, y_{1q})} \right)$$

$$= \frac{1}{2N^2}\mathrm{Re}\left( \sum_{m=1}^{2}\sum_{i=1}^{N}\sum_{j=1}^{N} \left( J_{m2}(x_{2i}, y_{2j}) - J_{m2}^{t}(x_{2i}, y_{2j}) \right)^{*} J_{m1}^{2}(x_{2i}, y_{2j}) f(x_{2i} - x_{1p}, y_{2j} - y_{1q}, z)\Delta x \Delta y \right)$$

$$+ \frac{1}{2N^2}\mathrm{Re}\left( \sum_{m=1}^{2}\sum_{i=1}^{N}\sum_{j=1}^{N} \left( J_{m1}(x_{2i}, y_{2j}) - J_{m1}^{t}(x_{2i}, y_{2j}) \right)^{*} J_{m2}^{2}(x_{2i}, y_{2j}) f(x_{2i} - x_{1p}, y_{2j} - y_{1q}, z)\Delta x \Delta y \right)$$

$$\frac{\partial L_{mase}}{\partial v_{12}^{1}(x_{1p}, y_{1q})} =$$

$$= \frac{1}{2N^2}\mathrm{Re}\left( i\sum_{m=1}^{2}\sum_{i=1}^{N}\sum_{j=1}^{N} \left( J_{m2}(x_{2i}, y_{2j}) - J_{m2}^{t}(x_{2i}, y_{2j}) \right)^{*} J_{m1}^{2}(x_{2i}, y_{2j}) f(x_{2i} - x_{1p}, y_{2j} - y_{1q}, z)\Delta x \Delta y \right)$$

$$+ \frac{1}{2N^2}\mathrm{Re}\left( i\sum_{m=1}^{2}\sum_{i=1}^{N}\sum_{j=1}^{N} \left( J_{m1}(x_{2i}, y_{2j}) - J_{m1}^{t}(x_{2i}, y_{2j}) \right)^{*} J_{m2}^{2}(x_{2i}, y_{2j}) f(x_{2i} - x_{1p}, y_{2j} - y_{1q}, z)\Delta x \Delta y \right)$$

Similarly, we have

$$\frac{\partial L_{mase}}{\partial u_{22}^{1}(x_{1p},y_{1q})}$$

$$=\frac{1}{2N^2}\mathrm{Re}\left(\sum_{m=1}^{2}\sum_{i=1}^{N}\sum_{j=1}^{N}\left(J_{m2}(x_{2i},y_{2j})-J_{m2}^{t}(x_{2i},y_{2j})\right)^{*}J_{m2}^{2}(x_{2i},y_{2j})f(x_{2i}-x_{1p},y_{2j}-y_{1q},z)\Delta x\Delta y\right)$$

$$\frac{\partial L_{mase}}{\partial v_{22}^{1}(x_{1p},y_{1q})}$$

$$=\frac{1}{2N^2}\mathrm{Re}\left(i\sum_{m=1}^{2}\sum_{i=1}^{N}\sum_{j=1}^{N}\left(J_{m2}(x_{2i},y_{2j})-J_{m2}^{t}(x_{2i},y_{2j})\right)^{*}J_{m2}^{2}(x_{2i},y_{2j})f(x_{2i}-x_{1p},y_{2j}-y_{1q},z)\Delta x\Delta y\right)$$

The gradients of $L_{mase}$ with respect to the components of the Jones matrix of the second layer $J^2(x_2,y_2)$ can be calculated directly as

$$\frac{\partial L_{mase}}{\partial u_{11}^{2}(x_{2p},y_{2q})}$$

$$=\frac{1}{2N^2}\mathrm{Re}\left(\sum_{m=1}^{2}\sum_{i=1}^{N}\sum_{j=1}^{N}\left(J_{1m}(x_{2p},y_{2q})-J_{1m}^{t}(x_{2p},y_{2q})\right)^{*}J_{1m}^{1}(x_{1i},y_{1j})f(x_{2p}-x_{1i},y_{2q}-y_{1j},z)\Delta x\Delta y\right)$$

$$\frac{\partial L_{mase}}{\partial v_{11}^{2}(x_{2p},y_{2q})}$$

$$=\frac{1}{2N^2}\mathrm{Re}\left(i\sum_{m=1}^{2}\sum_{i=1}^{N}\sum_{j=1}^{N}\left(J_{1m}(x_{2p},y_{2q})-J_{1m}^{t}(x_{2p},y_{2q})\right)^{*}J_{1m}^{1}(x_{1i},y_{1j})f(x_{2p}-x_{1i},y_{2q}-y_{1j},z)\Delta x\Delta y\right)$$

$$\frac{\partial L_{mase}}{\partial u_{12}^{2}(x_{2p},y_{2q})}$$

$$=\frac{1}{2N^2}\mathrm{Re}\left(\sum_{m=1}^{2}\sum_{i=1}^{N}\sum_{j=1}^{N}\left(J_{1m}(x_{2p},y_{2q})-J_{1m}^{t}(x_{2p},y_{2q})\right)^{*}J_{2m}^{1}(x_{1i},y_{1j})f(x_{2p}-x_{1i},y_{2q}-y_{1j},z)\Delta x\Delta y\right)$$

$$+\frac{1}{2N^2}\mathrm{Re}\left(\sum_{m=1}^{2}\sum_{i=1}^{N}\sum_{j=1}^{N}\left(J_{2m}(x_{2p},y_{2q})-J_{2m}^{t}(x_{2p},y_{2q})\right)^{*}J_{1m}^{1}(x_{1i},y_{1j})f(x_{2p}-x_{1i},y_{2q}-y_{1j},z)\Delta x\Delta y\right)$$

$$\frac{\partial L_{mase}}{\partial v_{12}^{2}(x_{2p},y_{2q})}=$$

$$=\frac{1}{2N^2}\mathrm{Re}\left(i\sum_{m=1}^{2}\sum_{i=1}^{N}\sum_{j=1}^{N}\left(J_{1m}(x_{2p},y_{2q})-J_{1m}^{t}(x_{2p},y_{2q})\right)^{*}J_{2m}^{1}(x_{1i},y_{1j})f(x_{2p}-x_{1i},y_{2q}-y_{1j},z)\Delta x\Delta y\right)$$

$$+\frac{1}{2N^2}\mathrm{Re}\left(i\sum_{m=1}^{2}\sum_{i=1}^{N}\sum_{j=1}^{N}\left(J_{2m}(x_{2i},y_{2j})-J_{2m}^{t}(x_{2i},y_{2j})\right)^{*}J_{1m}^{2}(x_{2i},y_{2j})f(x_{2i}-x_{1p},y_{2j}-y_{1q},z)\Delta x\Delta y\right)$$

$$\frac{\partial L_{mase}}{\partial u_{22}^2(x_{2p},y_{2q})}$$

$$=\frac{1}{2N^2}\mathrm{Re}\left(\sum_{m=1}^{2}\sum_{i=1}^{N}\sum_{j=1}^{N}\left(J_{2m}(x_{2p},y_{2q})-J_{2m}^{t}(x_{2p},y_{2q})\right)^{*}J_{2m}^{1}(x_{1i},y_{1j})f(x_{2p}-x_{1i},y_{2q}-y_{1j},z)\Delta x\Delta y\right)$$

$$\frac{\partial L_{mase}}{\partial v_{22}^2(x_{2p},y_{2q})}$$

$$=\frac{1}{2N^2}\mathrm{Re}\left(i\sum_{m=1}^{2}\sum_{i=1}^{N}\sum_{j=1}^{N}\left(J_{2m}(x_{2p},y_{2q})-J_{2m}^{t}(x_{2p},y_{2q})\right)^{*}J_{2m}^{1}(x_{1i},y_{1j})f(x_{2p}-x_{1i},y_{2q}-y_{1j},z)\Delta x\Delta y\right)$$

Besides the mean of the absolute squared error $L_{mase}$, we also add a boundary constraint loss $L_{bnd}$ into the total loss, i.e., $L=L_{mase}+L_{bnd}$. The definition and the gradient of the boundary loss are provided in the next supplementary section. Then we use an optimization algorithm L-BFGS method to minimize the total loss. For our case, this algorithm starts with an initial estimate of the input variables, i.e., all the components Jones matrixes of the two single layer metasurfaces, and proceeds iteratively to refine the input variables with a sequence of better estimates. The derivatives of the total loss are used as a key driver of the algorithm to identify the direction of steepest descent.

## S3. Precondition of Eq. 2 and boundary constraint in gradient descent optimization

The Jones matrixes of the two single layer metasurfaces are all symmetric. We consider a $2\times 2$ complex symmetric matrix *J*, and figure out the preconditions that it can be decomposed into the summation of two symmetric unitary matrixes as the following form:

$$J=\begin{bmatrix} a & b \\ b & c \end{bmatrix}=R(-\theta_1)\begin{bmatrix} e^{i\varphi_1} & 0 \\ 0 & e^{i\varphi_2} \end{bmatrix}R(\theta_1)+R(-\theta_2)\begin{bmatrix} e^{i\varphi_3} & 0 \\ 0 & e^{i\varphi_4} \end{bmatrix}R(\theta_2) \qquad \text{S3.1}$$

where $R(\theta)=\begin{bmatrix} \cos\theta & -\sin\theta \\ \sin\theta & \cos\theta \end{bmatrix}$ is the rotation matrix. It should be noted that the main results have been presented in the supplementary materials of reference[3], but missing some specific details, which will be discussed here.

**Theorem 1.** A $2\times 2$ symmetric unitary matrix $J$ can be factorized in the form $J=W\begin{bmatrix} e^{i\varphi_1} & 0 \\ 0 & e^{i\varphi_2} \end{bmatrix}W^{-1}$, where $W$ is unitary and can be chosen as real orthogonal matrix, i.e, $W=\begin{bmatrix} \cos\theta & -\sin\theta \\ \sin\theta & \cos\theta \end{bmatrix}$.

**Proof:** It is known that a unitary matrix is unitarily diagonalizable and its eigenvalues are unimodular, therefore $J=W\Lambda W^{-1}$, $\Lambda=\begin{bmatrix} e^{i\varphi_1} & 0 \\ 0 & e^{i\varphi_2} \end{bmatrix}$ and $WW^{\dagger}=I$. Here $W^{\dagger}$ is the conjugate transpose of $W$.

Since $J$ is also symmetric, there exist real symmetric matrices $A$ and $B$ such that $J=A+iB$. Then $J^{\dagger}J=(A-iB)(A+iB)=A^2+B^2+i(AB-BA)=I$, which imply that $A$ and $B$ commute $AB=BA$. As real symmetric matrix can have real eigenvectors, a pair of real commuting symmetric matrices can be simultaneously diagonalized by the same set of real eigenvectors, which are also the eigenvectors of $J$. Therefore, $W$ can be real matrix. Also, a general expression of a 2 × 2 unitary matrix can be written in the form:

$$W=e^{i\alpha/2}\begin{bmatrix} e^{i\alpha_1}\cos\theta & e^{i\alpha_2}\sin\theta \\ -e^{-i\alpha_2}\sin\theta & e^{-i\alpha_1}\cos\theta \end{bmatrix} \qquad \text{S3.2}$$

When it is real, it becomes $W = \begin{bmatrix} \cos\theta & -\sin\theta \\ \sin\theta & \cos\theta \end{bmatrix}$.

**Theorem 2.** A $2\times 2$ symmetric matrix *J* can be decomposed into the summation of two symmetric unitary matrix if the singular values of *J* are all less than or equal to 2.

**Proof:** With singular value decomposition, one can factorize any arbitrary $2\times 2$ matrices *J* in the form

$$J = W\Sigma V^{\dagger} \qquad \text{S3.3}$$

where $W$ and $V$ are both unitary matrix, and $\Sigma$ is a real diagonal matrix, $\Sigma = \begin{bmatrix} r_1 & 0 \\ 0 & r_2 \end{bmatrix}$, $r_1$ and $r_2$ are nonnegative real numbers and called as the singular values of *J*.

The columns of *W* (left-singular vectors) and *V* (right-singular vectors) are the eigenvectors of $JJ^{\dagger}$ and $J^{\dagger}J$, respectively. Assume that $J^{\dagger}J\mathbf{x}_1 = \lambda\mathbf{x}_1$ and $JJ^{\dagger}\mathbf{x}_2 = \lambda\mathbf{x}_2$. If *J* is symmetric, we have $J^{\dagger} = \overline{J}^{T} = \overline{J}$, where $J^{T}$ denotes the matrix transpose of *J* and $\overline{J}$ is the conjugate of *J*, then $J^{\dagger}J\mathbf{x}_1 = \overline{J}J\mathbf{x}_1 = \lambda\mathbf{x}_1$ and $JJ^{\dagger}\mathbf{x}_2 = J\overline{J}\mathbf{x}_2 = \lambda\mathbf{x}_2$. So $\overline{J}J\,\overline{\mathbf{x}}_2 = \lambda\overline{\mathbf{x}}_2$. Therefore, we can choose $\mathbf{x}_1 = \overline{\mathbf{x}}_2$, then $W = \overline{V}$. Indeed, for symmetric matrices, Takagi decomposition is a special case of the singular value decomposition. Takagi tell us there is a unitary *W* such that $J = W\Sigma W^{T}$ if $J$ is a symmetric matrix, Again, we have $V^{\dagger} = W^{T}$, i.e., $\overline{V} = W$.

It can be proved that any complex number *r* can be decomposed as two complex numbers with unit amplitude when $|r| \le 2$. Therefore, if the singular values of *J* are all less or equal to 2, i.e., $r_1, r_2 \le 2$, we have

$$J = W\Sigma V^{\dagger} = W\Sigma W^T = W\begin{bmatrix} e^{i\varphi_1} + e^{i\varphi_3} & 0 \\ 0 & e^{i\varphi_2} + e^{i\varphi_4} \end{bmatrix} W^T$$

$$= W\begin{bmatrix} e^{i\varphi_1} & 0 \\ 0 & e^{i\varphi_2} \end{bmatrix} W^T + W\begin{bmatrix} e^{i\varphi_3} & 0 \\ 0 & e^{i\varphi_4} \end{bmatrix} W^T \qquad \text{S3.4}$$

As $W$ is unitary, the two terms $W\begin{bmatrix} e^{i\varphi_1} & 0 \\ 0 & e^{i\varphi_2} \end{bmatrix} W^T$ and $W\begin{bmatrix} e^{i\varphi_3} & 0 \\ 0 & e^{i\varphi_4} \end{bmatrix} W^T$ are both symmetric unitary matrixes, too. According to Theorem 1, they must can be factorized in the form $R(-\theta_1)\begin{bmatrix} e^{i\varphi_1} & 0 \\ 0 & e^{i\varphi_2} \end{bmatrix} R(\theta_1)$.

According to Theorems 1 and 2, the precondition for the validity of Eq. S3.1 (i.e., Eq. 2 in the main text) is that the singular values of the symmetric Jones matrix $J$ are both no more than 2. The singular values ($r_1$ and $r_2$) of $J$ are the square roots of eigenvalues ($\lambda$) from $J^{\dagger}J$, which can be solved by

$$\det\left(J^{\dagger}J\text{-}\lambda I\right) = \lambda^2 - \left(|a|^2 + 2|b|^2 + |c|^2\right)\lambda + \left|ac - b^2\right|^2 = 0 \qquad \text{S3.5}$$

To ensure $r_1, r_2 \le 2$, both the two eigenvalues $\lambda_1, \lambda_2 \le 4$. Then we have

$$\begin{cases} 4^2 - 4\left(|a|^2 + 2|b|^2 + |c|^2\right) + \left|ac - b^2\right|^2 \ge 0 & \text{S3.6.1} \\ \left(|a|^2 + 2|b|^2 + |c|^2\right)/2 \le 4 & \text{S3.6.2} \end{cases}$$

We define $f(a,b,c) = 4^2 - 4\left(|a|^2 + 2|b|^2 + |c|^2\right) + \left|ac - b^2\right|^2$, and its minimum value is

$$f(a,b,c)_{\min} = 4^2 - 4\left(|a|^2 + 2|b|^2 + |c|^2\right) + \left(|a||c| - |b|^2\right)^2$$

$$= \left(|b|^2 - |a||c| - 4\right)^2 - 4\left(|a| + |c|\right)^2$$

$$= \left(|b|^2 - \left(|a| - 2\right)\left(|c| - 2\right)\right)\left(|b|^2 - \left(|a| + 2\right)\left(|c| + 2\right)\right) \ge 0 \qquad \text{S3.7}$$

According to Eq. S3.6.2, $\left(|b|^2 - \left(|a| + 2\right)\left(|c| + 2\right)\right) \le 0$, then we have

$$\begin{cases} \left(|b|^2 - (|a|-2)(|c|-2)\right) \le 0 & \text{S3.8.1} \\ \left(|a|^2 + 2|b|^2 + |c|^2\right)/2 \le 4 & \text{S3.8.2} \end{cases}$$

One simple sufficient condition for the above two inequations is

$$\begin{cases} |a| + |b| \le 2 \\ |b| + |c| \le 2 \end{cases} \qquad \text{S3.9}$$

For our designed two layer metasurfaces, their Jones matrix components are confined by defining the following boundary loss:

$$\begin{aligned} L_{bnd} = &\operatorname{Re}LU\left(\left|J_{11}^1\right| + \left|J_{12}^1\right| - 2\right) + \operatorname{Re}LU\left(\left|J_{22}^1\right| + \left|J_{12}^1\right| - 2\right) \\ &+ \operatorname{Re}LU\left(\left|J_{11}^2\right| + \left|J_{12}^2\right| - 2\right) + \operatorname{Re}LU\left(\left|J_{22}^2\right| + \left|J_{12}^2\right| - 2\right) \end{aligned} \qquad \text{S3.10}$$

where $ReLU$ is the rectified linear unit. For numerical calculation, the metasurface layer is divided into pixels and the loss is then averaged over all pixels. The gradients of this boundary loss with respect to the Jones matrix components of the two single layers can be readily obtained.

## S4. Design of the target equivalent Jones matrix and comparison with optimized results

The designed target equivalent Jones matrix has four components with each having two degrees of freedoms (DOFs), amplitude and phase. Four nanoprinting images (weather symbol images, three intensity levels) and four holographic images (letter strings “XX”, “XY”, “YX” and “YY”) are encoded into the four components of the equivalent Jones matrix to serve as the targets. The four channels are independent, and for each of them, we use a modified Gerchberg-Saxton algorithm to obtain its amplitude and phase distributions as the same as our previous work[4]. The amplitude distribution of the equivalent Jones matrix is chosen as that of the input nanoprinting image. Then a random phase distribution is added to it and propagated to the holographic image plane using Rayleigh–Sommerfeld diffraction method that propagates in silicon dioxide substrate with refractive index of 1.45. The amplitude of the holographic image is then substituted with the designed one, and propagated backward to the nanoprinting plane. At the nanoprinting plane, the amplitude constraint is also taken. After several iterations, when the computed holographic amplitude at the far-field holographic plane is close to the target ones, we can obtain the phase distribution of the equivalent Jones matrix.

The operating wavelength is designed at 808 nm and the sampling period is chosen as 800 nm to avoid high order diffractions. The design nanoprintings have $320 \times 320$ pixels with a total size of $256\mu m \times 256\mu m$. The holographic images are designed at 2500 μm above the nanoprinting in glass substrate, corresponding to a maximal numerical aperture about $n \sin\theta \approx 1.45 \times 0.1 = 0.145$. Usually, a smaller numerical aperture indicates a slower fluctuation of the phase distribution, i.e., smaller phase difference between adjacent pixels. We choose such a long nanoprinting-holography distance to ensure that the small phase fluctuation do not strongly affect the observation of nanoprinting images.

Considering the convenience in the experimental measurement, we set the vertical distance between the two single layer metasurfaces as 150 μm. The material between the two layers is air with refractive index 1.0. The targets of the amplitude and phase distributions of the four components of the equivalent Jones matrix are shown in Figure S2, where the optimized results based on the gradient descent optimization method are also shown. It can be observed that the phase distributions between the target and optimized results are almost identical, while the amplitude distributions differ a little bit around the boundaries. The reason is that the amplitude at the boundary has a sharp fluctuation including high angular spectrum frequencies which is beyond our design (Note that our design has a numerical aperture of 0.145). From this aspect, we can see that the optimized results agree very well with the target ones, demonstrating the feasibility of the gradient descent optimization to obtain the arbitrary equivalent Jones matrix.

The optimal input values, i.e., the Jones matrixes of the two single layers are provided in Figure S3. The amplitudes of all components do not exceed over 2 due to the introducing of the boundary constraint loss.

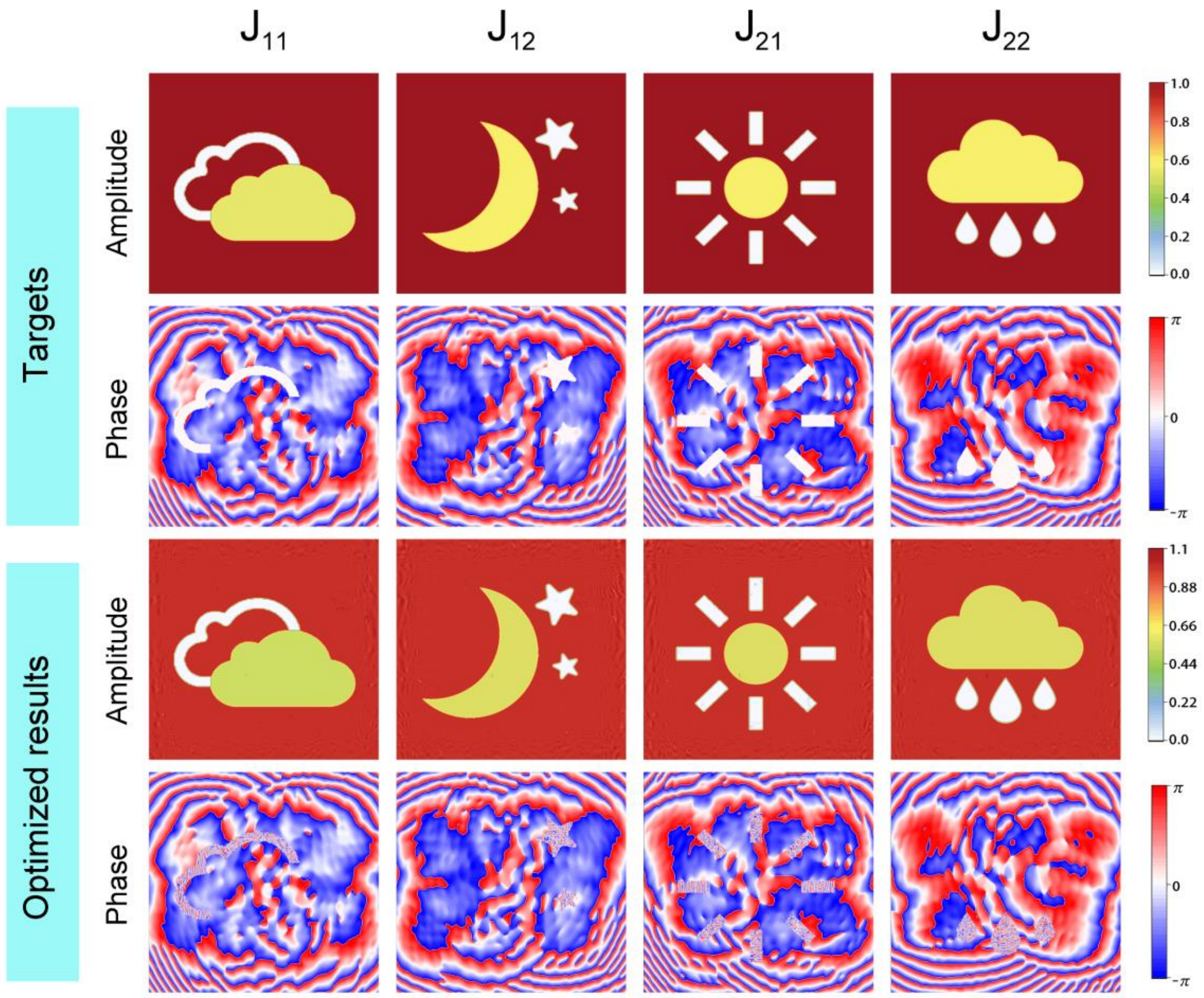

Figure S2. The comparisons of the four components of the equivalent Jones matrix between the targets and optimized results.

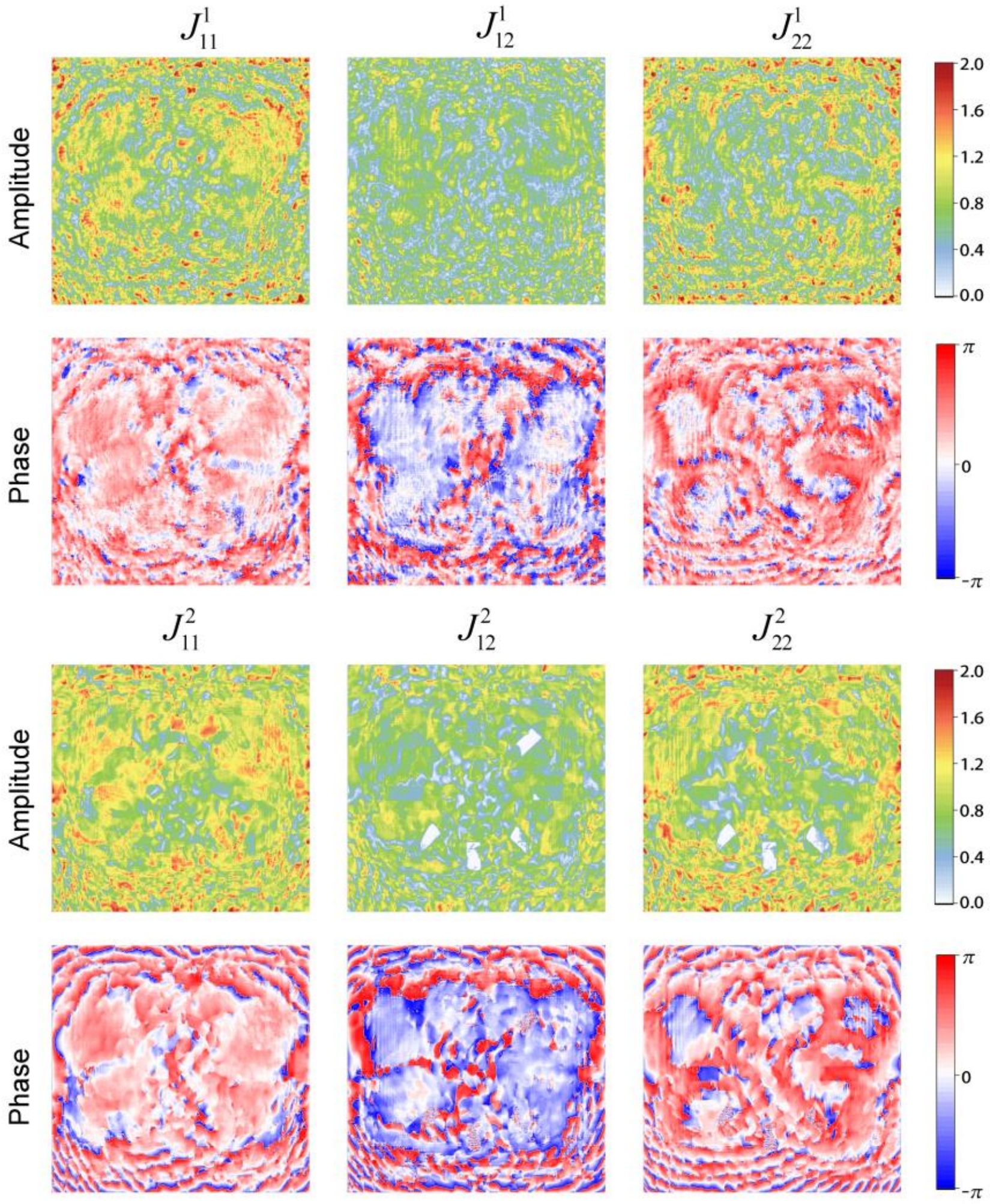

Figure S3. Optimized input variables of the Jones matrix of the two single layers from the gradient descent algorithm. The non-diagonal entries of the Jones matrix are the same due to the symmetry of single layer and therefore we only present the 12th component here.

## S5. Metasurface unit design

As outlined in the main text, the symmetric Jones matrix of the single layer with six DOFs can be constructed by two nano pillars in one unit. The nano pillars are made of crystal-silicon with a fixed height of 600 nm on glass substrate. A schematic view of such nano pillar is shown in Figure S4 (a). Due to the symmetry, the unit cell support two propagation modes along its *x* and *y* axis. The transmission magnitude and phase shift introduced by the nano pillar with *x*-polarized and *y* polarized incidences, as a function of the transverse dimensions of the nano pillars d*x* and d*y*, are shown in Figure S4 (b-e). The results are obtained numerically via FDTD simulations. Note that the two layer metasurfaces are arranged front-to-front to maintain a homogeneous air environment between them. The transmission properties of nanopillars of the two layers (one is from substrate to air and the other is opposite) are the same due to reciprocity principle and can be extracted from the same database.

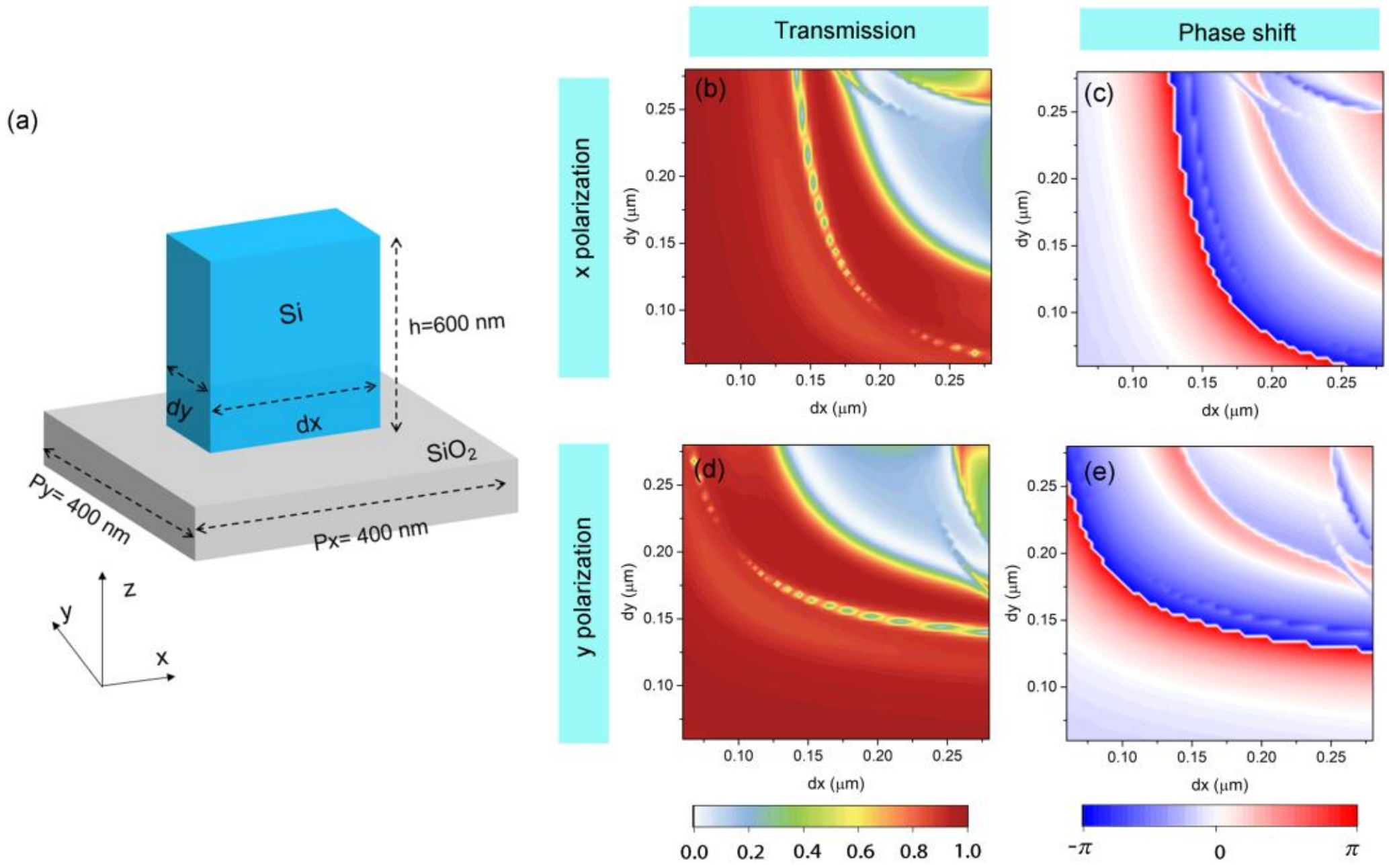

Figure S4: Metasurface unit design. (a) Schematic of the metasurface unit cell composed of a rectangle nanopillar made of silicon (Si) standing on a glass substrate. The period along *x* and *y* direction is 400 nm, and the height is fixed of 600 nm. (b-c) Normalized transmission (b) and phase response (b, in radians) of the nanopillar at incident wavelength 808 nm as a function of the transverse dimensions, dx and dy, for incident x-polarization. (d-e) Normalized transmission (d) and phase response (e, in radians) of the nanopillar at incident wavelength 808 nm as a function of the transverse dimensions, dx and dy, for incident y-polarization. The optical response for y-polarization response is obtained by swapping x and y of x-polarization response due to the nature of symmetry.

For a given Jones matrix with six DOFs, we can extract the rotational angles and the phase shifts along *x* and *y* directions of both nano pillars, according to Eq. S3.1. For each individual nano pillar, the transverse dimensions are chosen as follows[5]: (1) first set a pre-defined average transmission magnitude $t_{avg}$. This value is basically determined by the overall transmissions of the nano pillar (Figure S4b) with different transverse dimensions and should not be small. In our work, we choose $t_{avg} = 0.98$. (2) calculate the complex-valued errors $\varepsilon_x = \left| t_{avg} e^{i\varphi_{x,desired}} - t_{simulated} e^{i\varphi_{x,simulated}} \right|$ and $\varepsilon_x = \left| t_{avg} e^{i\varphi_{x,desired}} - t_{simulated} e^{i\varphi_{x,simulated}} \right|$, and choose the large one $\varepsilon_{\max} = \max(\varepsilon_x, \varepsilon_y)$. (c) determine the configuration that minimizes $\varepsilon_{\max}$ for all possible dimensions.

## S6. FDTD simulation of the realistic structures with different optical setups

The numerical calculations based on the diffraction theory in Figure S2 show good results. However, for realistic structures, many factors, such as the errors between the designed phase shifts and real ones of the nano pillars, the residual zero order diffractions and the coupling between nano pillars may all affect the optical performances. Therefore, it is necessary to perform full wave electromagnetic simulations (FDTD simulations) with the realistic structures, which can also provide a guidance for design and measurement. The structure in our work consists of two layer metasurfaces, with a side length of 256 μm and a vertical distance of 150 μm. Obviously, it is unrealistic to simulate the whole structure and thus we turn to a scaled-down version with $120\times120$ pixels (96 μm $\times$96 μm) and gap distance of 100 μm. The holographic images are designed at a shorter distance of 400 μm above the second layer.

It is still impossible to simulate the whole structure as the dimensions along the three directions are all around hundreds of micrometers. Since the average transmission of all the nanopillars is about 0.98, it is reasonable to neglect the reflections between the two layers. Therefore, the simulation of the whole structure can be divided into three steps: (1) perform the simulation with only the first layer and obtain the near fields at the output plane of the first layer; (2) The equivalent electric and magnetic currents obtained from above near fields are used to calculate the far fields at the incident plane of the second layer, which are then imported as source to excite the second layer. (3) perform the simulation of the second layer with the imported source and obtain the near fields at the output plane of the second layer. With the above treatment, we decompose the simulation of the whole structure into two simulations with a shorter size along the *z* direction. Each simulation can be performed within an acceptable time (about 10 hours on a workstation with CPU AMD 3990x and RAM 256G).

The obtained near fields at the output plane of the second layer are used to extract the nanoprinting images and holographic images with different numerical aperture centers and sizes. Assuming that the far fields in the angular spectrum domain are $A(u,v,0)$, the images at distance *h* above the second layer with numerical aperture size *NA* at center $(u_0,v_0)$ can be calculated as

$$H(x_2,y_2)=\iint_{u,v} A(u,v,0)\cdot G(u,v,h)\cdot F(u,v)\cdot \exp(i2\pi(ux_2+vy_2))dudv \quad \text{(S6.1)}$$

where $G(u,v,h)=\exp(i2\pi h(\lambda^{-2}-u^2-v^2)^{1/2})$ and $F(u,v)$ is the filter in Fourier plane, as

$$F(u,v)=\begin{cases}1 & if\ \sqrt{(u-u_0)^2+(v-v_0)^2}\le\dfrac{NA}{\lambda}\\ 0 & else\end{cases} \quad \text{(S6.2)}$$

The nanoprinting and holographic images can be obtained by set *h*=0 and *h*=400 μm. We implement different optical setups: normal incidence-normal detection, oblique incidence-normal detection, and oblique incidence-oblique detection. The observation numerical aperture is chosen *NA*=0.2 and the center of the oblique detection is set $(u_0,v_0)=(\frac{0.3}{\lambda},0)$. The full simulated results of the three cases are shown in Figure S5.

It can be observed that the nanoprinting images are strongly affected by the different optical setups. This is because the designed nanoprintings and zeros orders are on the same order of magnitude. When applying the normal incidence normal detection setup, both the zero order diffractions from two layers are detected, which results in a very poor image quality of nanoprinting. By applying the oblique incidence normal detection, only the zero order diffraction from the second layer are detected. The nanoprinting for the oblique incidence oblique detection setup has the highest image quality among the three setups since the zero order diffractions from both layers are filtered.

On the other hand, the holographic images for all cases can be observed clearly and have similar image fidelities. This is because the magnitudes of the holographic images in design are much higher than that of zero orders, which thus do not strongly affect the holographic images. Due to the limit of computing capacity, the input images in the simulation have a small size and a low resolution. We anticipate that the image qualities can be increased for practical structures that have much larger number of pixels.

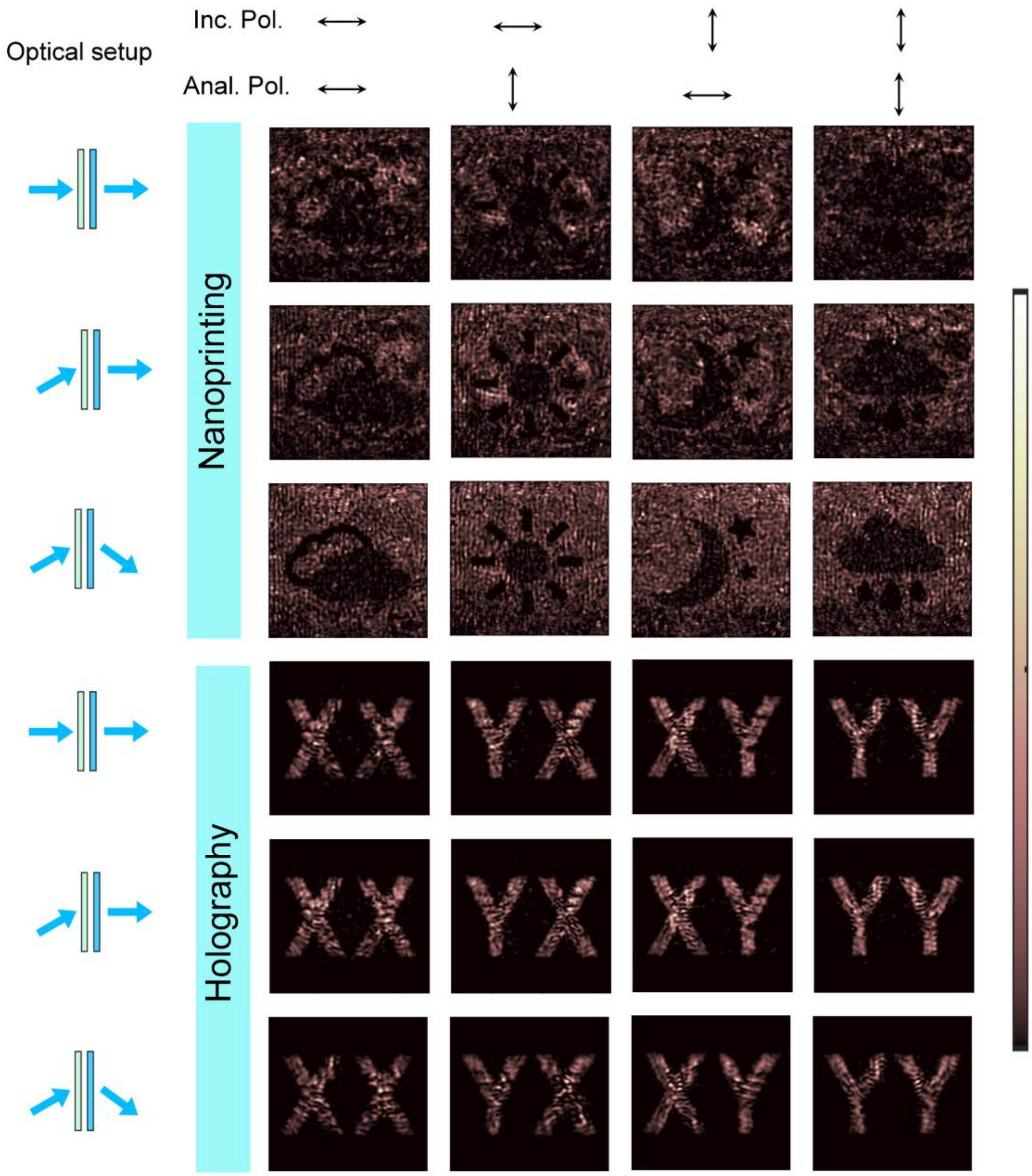

Figure S5. FDTD simulated results of the nanoprintings and holographic images with different optical setups. The schematic views of different optical setups are shown in the left column. The incident and analyzed polarizations are indicated at the top two rows.

## S7. Effects of detour phase in multi-element unit design

Figure S6a presents a metasurface unit cell with one element, which imposes amplitude modulation $A$ and phase shift $\varphi$ on the incident light. If one expects to realize the same optical response by two-element unit design (Figure S6b), a simple way is to use the same two elements as that in Figure S6a. However, this is not always the best choose. For oblique incidence and oblique scattering, the lateral displacements of the two elements introduce detour phases within this unit cell. Assuming that the two elements are evenly distributed that are located at $x$ coordinates of -$P$/4 and $P$/4, where $P$ is the period, the summation of the scatterings from the two elements are $Ae^{i\varphi}(e^{i\Delta\varphi}+e^{-i\Delta\varphi})=2A\cos\Delta\varphi e^{i\varphi}$, where $\Delta\varphi=\frac{2\pi}{\lambda}\frac{P}{4}(\sin\theta_1+\sin\theta_2)$, $\theta_1$ is the incident angle and $\theta_2$ is the scattering angle.

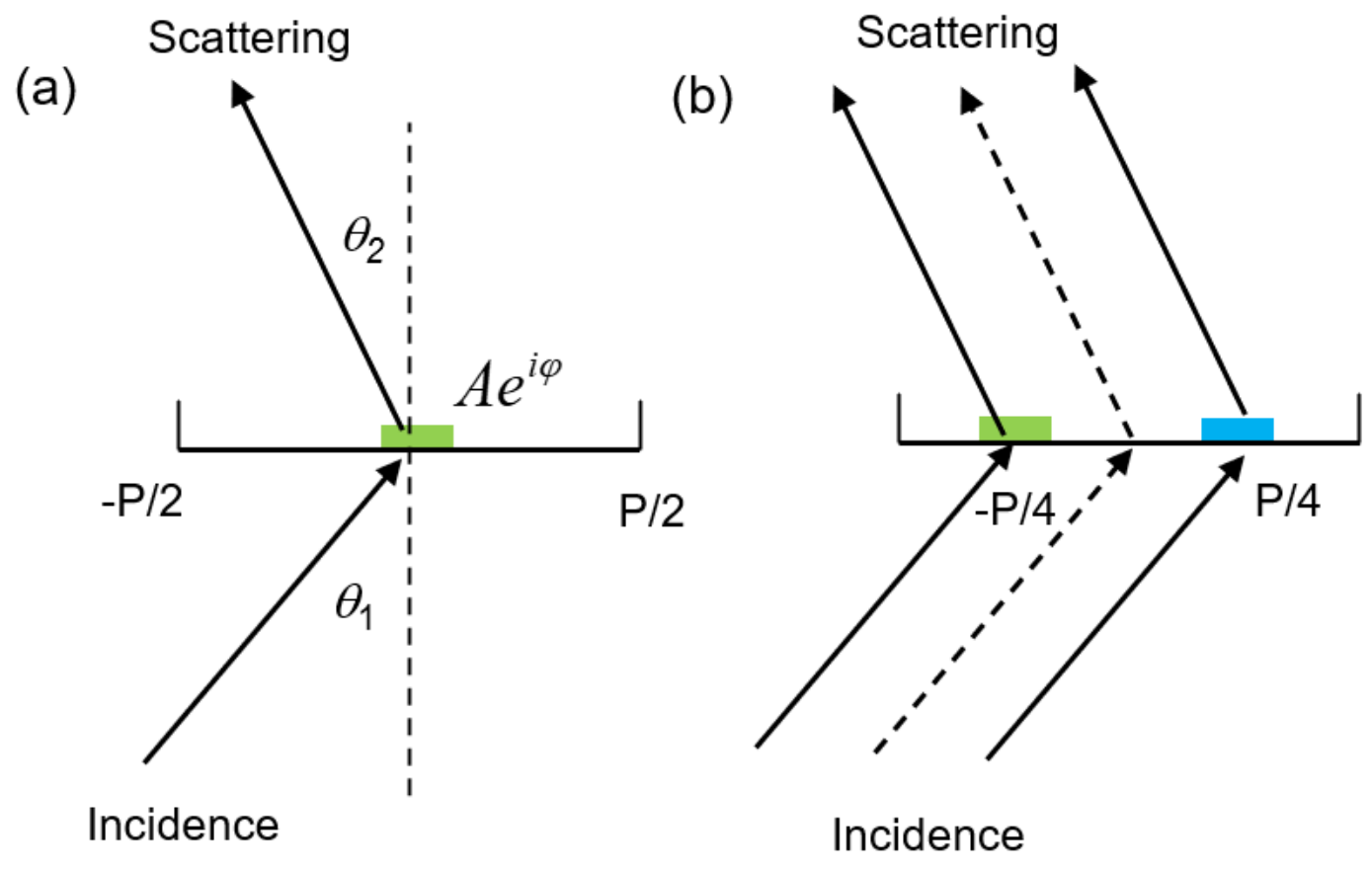

Figure S6. Schematics of the metasurface unit cell with one element and two elements.

It is shown that the duplication of the element in the unit cell do not alter the phase shift (must be evenly distributed) but impose an amplitude modulation of $2A\cos\Delta\varphi$. To satisfy the Shannon–Nyquist sampling theorem, the scattering angle is limited by $\sin\theta_2\leq\lambda/2P$. At the limit scattering angle, $\Delta\varphi=\pi/4$ and the efficiency decreases to $\cos^2\Delta\varphi=50\%$. It is also obvious that the efficiency will decrease for oblique incidence. If we consider the detour phase and individually design the two elements, the total scattering becomes $Ae^{i(\varphi_1+\Delta\varphi)}+Ae^{i(\varphi_2-\Delta\varphi)}$, where $\varphi_1$ and $\varphi_2$ are the phase shifts of the two elements. The best optical performance occurs when $\varphi_1=\varphi-\Delta\varphi$ and $\varphi_2=\varphi+\Delta\varphi$, and the total optical field is $2Ae^{i\varphi}$.

As for the unit design in Figure 2c of main text, the above two elements correspond to the nano pillars *AB* and *A*'*B*'. For each pixel of the single layer, a special amplitude modulation and phase shift is designed for each component of its Jones matrix. In our work, the Jones matrix of the nano pillars *AB* and *A*'*B*' are individually designed for oblique incidence (metasurface 1, Figure 2d) or oblique scattering (metasurface 2, Figure 2d) according to the above analysis. A comparison of the FDTD simulations between considering the detour phase (individually designing nano pillars *AB* and *A*'*B*') and without (*A*'*B*' is a duplication of *AB*) is shown in Figure S7. Clearly, the nanoprinting images with the consideration of the detour phase have much higher fidelities than that without the consideration of the detour phase.

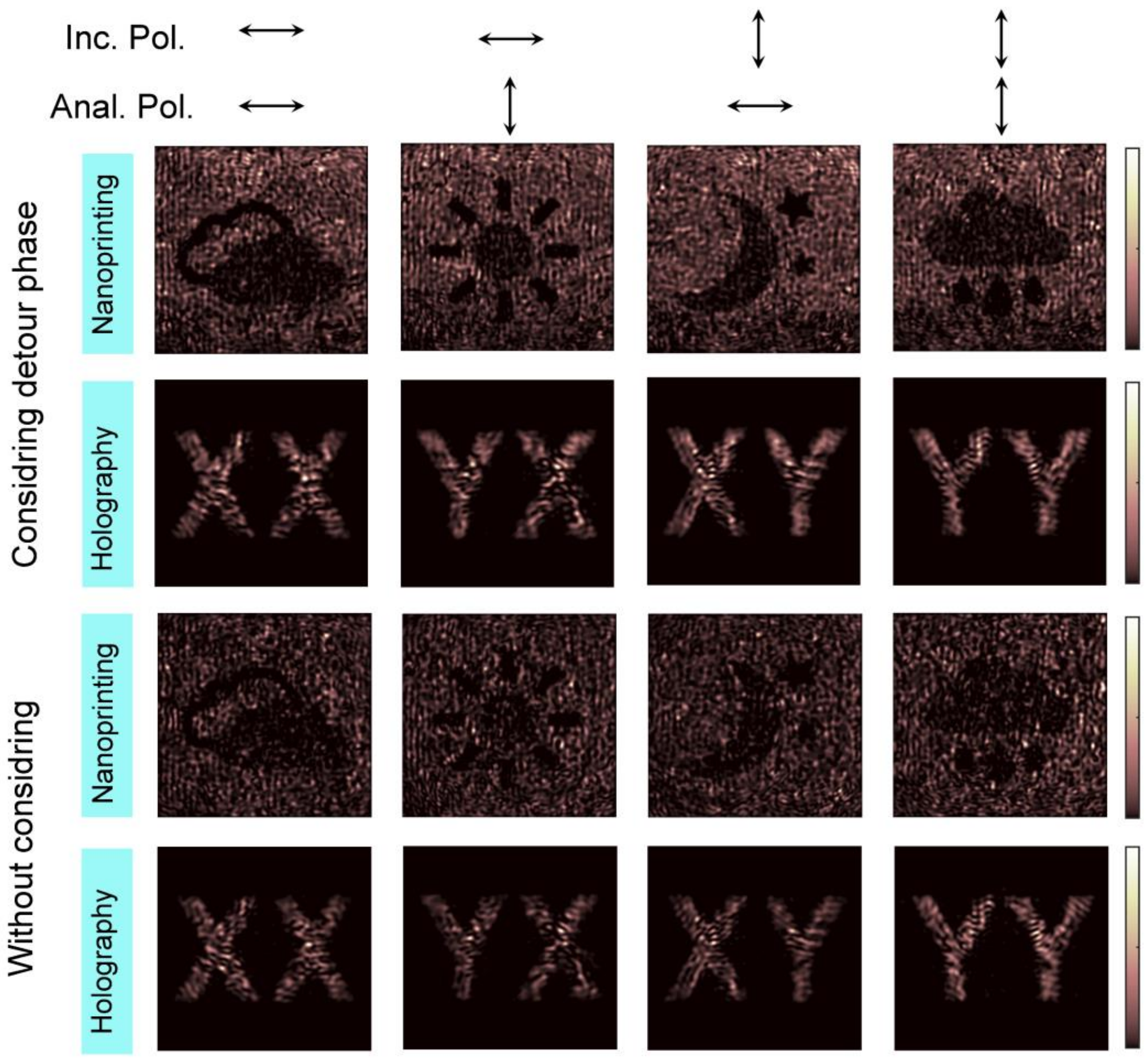

Figure S7. FDTD simulations of the nanoprinting and holographic images with the consideration of the detour phase in the unit pixel (first and second rows) and without (third and fourth rows).

## S8. Alignment sensitivity of the two layer metasurfaces

The optical performance of nanoprinting and holographic images are dependent on the alignment of the two layer metasurfaces. Figure S8 shows the simulated results of the images with different translational in-plane $x$-shift values. Basically, the images can be distinctly observed within a shift value of 5 μm. In experiment, the two metasurfaces are mounted on two 3D translational stages (Thorlabs, MBT616D/M), which can provide a resolution about 100~200 nm, thus satisfying our alignment requirement.

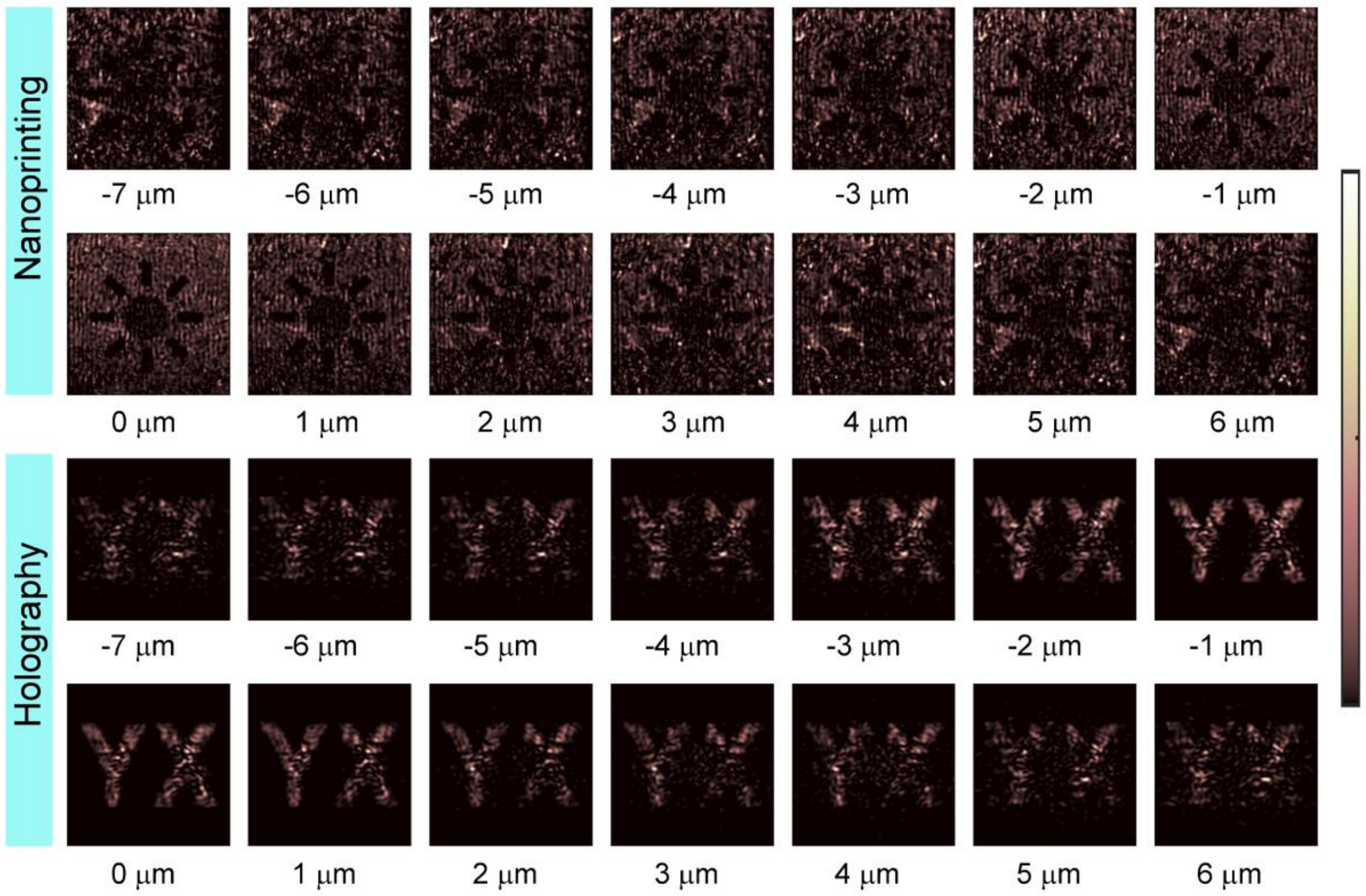

Figure S8. Simulated nanoprinting and holographic images with x-polarized incidence and y-polarized analyzation for different translational shift values.

## S9. Optical setup for measurement

A schematic view of our optical setup with detailed parameters is shown in Figure S9. The focal lengths of the objective (RMS20X-PF, 20X, N.A.=0.5, Thorlabs Inc.), Lens 1, Lens 2 and Lens 3 are $f_o$=9 mm, $f_1$=100 mm, $f_2$=200 mm and $f_3$=200 mm, respectively. A beam with diffraction angle $\theta$ firstly focuses at position of $d=f_o\sin\theta$ away from the center axis at the back Fourier plane of the objective. After passing through Lens 1 and Lens 2, it focuses at the back focal plane of Lens 2 with a distance of $D= f_2 / f_1 * d = f_2 f_o \sin\theta / f_1$ away from the center axis.

For optical measurement, we apply a phase shift gradient of $k_x/k$=0.3 on the second layer, corresponding to $\sin\theta$=0.3. Therefore, the center of the filter is located at $D$=5.4 mm away from the center axis. The aperture size $a$ of the filter is chosen to have a numerical aperture of N. A.= 0.2, corresponding to a diameter of $a=2f_2 f_o$ N.A. $/f_1$ =7.2 mm. In measurement, a continuously variable iris is used for the filter.

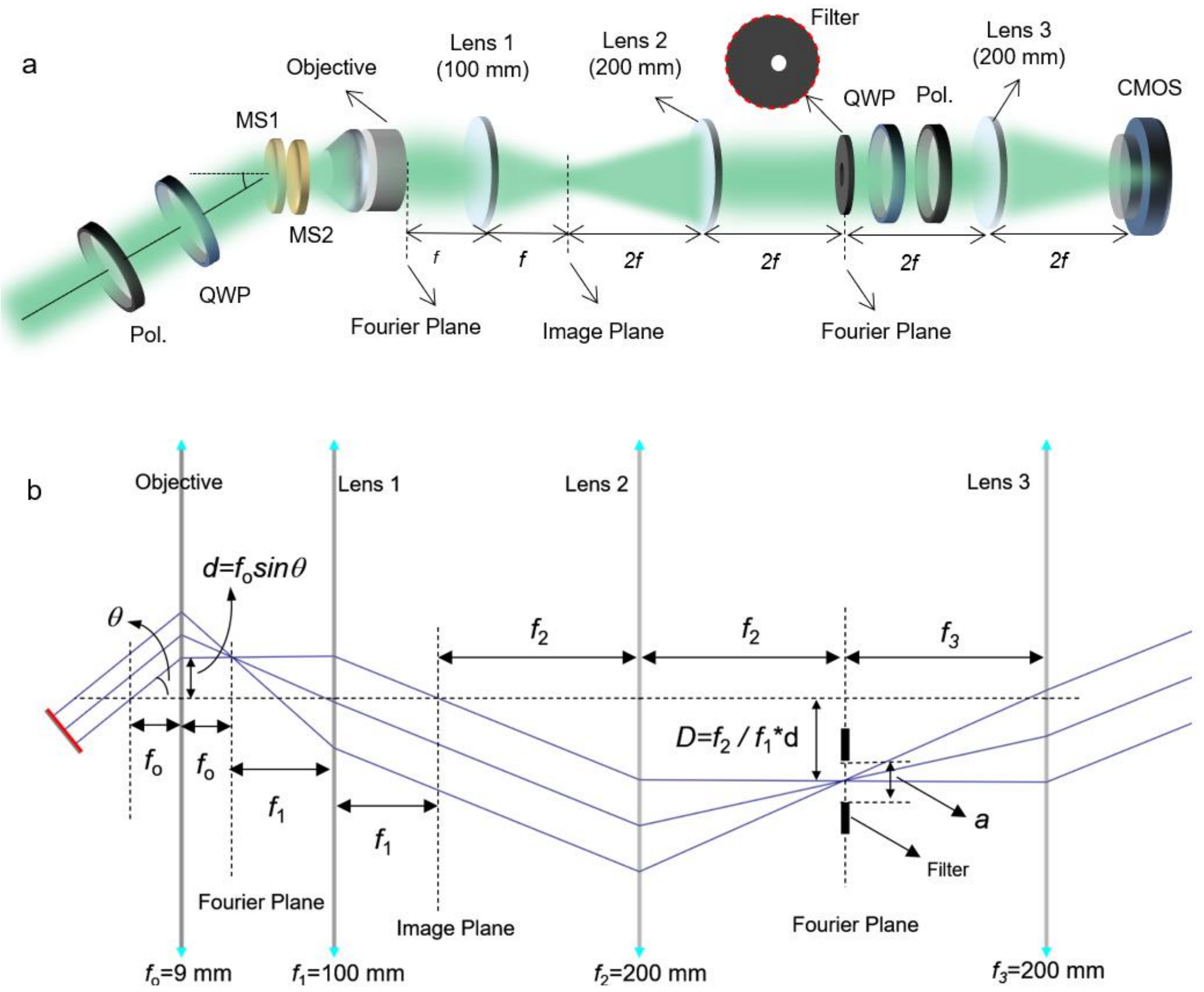

Figure S9. a, Schematic of the optical setup for measurements; b, Schematic of the propagation of a plane wave at oblique incidence, showing the detailed parameters for spatial filtering.

## S10. Calculation of optical responses with rotation between the two layer metasurfaces

Here, we investigate the optical responses of the two layer system with different rotation angle. The first layer is assumed to be fixed, and the second layer is rotated anticlockwise with an angle $\phi$. The Jones matrix components of the second layer after rotation are given by:

$$\begin{bmatrix} J_{11}^{2\phi}(x_2,y_2), J_{12}^{2\phi}(x_2,y_2) \\ J_{21}^{2\phi}(x_2,y_2), J_{22}^{2\phi}(x_2,y_2) \end{bmatrix} = \begin{bmatrix} J_{11}^{2}(x_2,y_2), J_{12}^{2}(x_2,y_2) \\ J_{21}^{2}(x_2,y_2), J_{22}^{2}(x_2,y_2) \end{bmatrix} \tag{S10.1}$$

for $\phi = 0°$,

$$\begin{bmatrix} J_{11}^{2\phi}(x_2,y_2), J_{12}^{2\phi}(x_2,y_2) \\ J_{21}^{2\phi}(x_2,y_2), J_{22}^{2\phi}(x_2,y_2) \end{bmatrix} = \begin{bmatrix} J_{22}^{2}(-y_2,x_2), -J_{21}^{2}(-y_2,x_2) \\ -J_{12}^{2}(-y_2,x_2), J_{11}^{2}(-y_2,x_2) \end{bmatrix} \tag{S10.2}$$

for $\phi = 90°$,

$$\begin{bmatrix} J_{11}^{2\phi}(x_2,y_2), J_{12}^{2\phi}(x_2,y_2) \\ J_{21}^{2\phi}(x_2,y_2), J_{22}^{2\phi}(x_2,y_2) \end{bmatrix} = \begin{bmatrix} J_{11}^{2}(-x_2,-y_2), J_{12}^{2}(-x_2,-y_2) \\ J_{21}^{2}(-x_2,-y_2), J_{22}^{2}(-x_2,-y_2) \end{bmatrix} \tag{S10.3}$$

for $\phi = 180°$,

$$\begin{bmatrix} J_{11}^{2\phi}(x_2,y_2), J_{12}^{2\phi}(x_2,y_2) \\ J_{21}^{2\phi}(x_2,y_2), J_{22}^{2\phi}(x_2,y_2) \end{bmatrix} = \begin{bmatrix} J_{22}^{2}(y_2,-x_2), -J_{21}^{2}(y_2,-x_2) \\ -J_{12}^{2}(y_2,-x_2), J_{11}^{2}(y_2,-x_2) \end{bmatrix} \tag{S10.4}$$

for $\phi = 270°$, where the superscript $\phi$ indicates the rotation case, and the Jones matrix components at the right hand side of the above formulas are the ones without rotation, i.e., the initial input variables in the gradient descent optimization algorithm. Note that both the Jones matrix component values and the coordinates are changed after rotation. According to Eq. S1.5, the equivalent Jones matrix after rotation is given by

$$J_{mn}^{\phi}(x_2,y_2) = \sum_{q=1,2} J_{mq}^{2\phi}(x_2,y_2) \iint_{x_1,y_1} J_{qn}^{1}(x_1,y_1) \cdot f(x_2-x_1, y_2-y_1, z) dx_1 dy_1 \tag{S10.5}$$

The holographic image with a distance $h$ above the second layer is

$$H_{mn\phi}(x_3,y_3) = \iint_{x_2,y_2} J_{mn}^{\phi}(x_2,y_2) \cdot f(x_3-x_2, y_3-y_2, h) dx_2 dy_2 \tag{S10.6}$$

where $m$, $n$=1 indicates the $x$-polarized incidence or analyzation, $m$, $n$=2 indicates $y$-polarized incidence or analyzation. Therefore, there are 16 cases by combining the four rotation angles and the four incidence-analyzation polarizations.

For numerical calculation, all the planes are sampled to $N \times N$ equidistant grids with sampling intervals of $\Delta x = \Delta y = P$. We define a loss of

$$L = \frac{1}{16N^2} \sum_{\phi=1}^{4} \sum_{m=1}^{2} \sum_{n=1}^{2} \sum_{i=1}^{N} \sum_{j=1}^{N} \left( \left| H_{mn\phi}(x_{3i}, y_{3j}) \right| - \left| H_{mn\phi}^{t}(x_{3i}, y_{3j}) \right| \right)^2 \qquad \text{(S10.7)}$$

where we use the index $\phi = 1, 2, 3, 4$ to represent the cases with rotational angle of $\phi$= 0°,90°, 180°, 270°, $H_{mn\phi}^{t}$ indicates the designed target holographic images under $mn$ incidence-analyzation polarizations and rotational angle $\phi$. Such definition of the loss does not impose any constraints on the phases of the predicted holographic images. The boundary constraint loss is also added in the total loss. The gradient calculations of this loss are more complicated than that of the EQJM, but the basic ideas are the same, the details of which are not present here anymore. Note that there is a freedom of the magnitude ratio for the target images, and we set the maximal amplitude value as 2 for the balance consideration of the astringency and the efficiency.

We input 16 holographic images as the targets, and the optimized results of the gradient descent optimization are shown in Fig. S10, which agrees well with our targets.

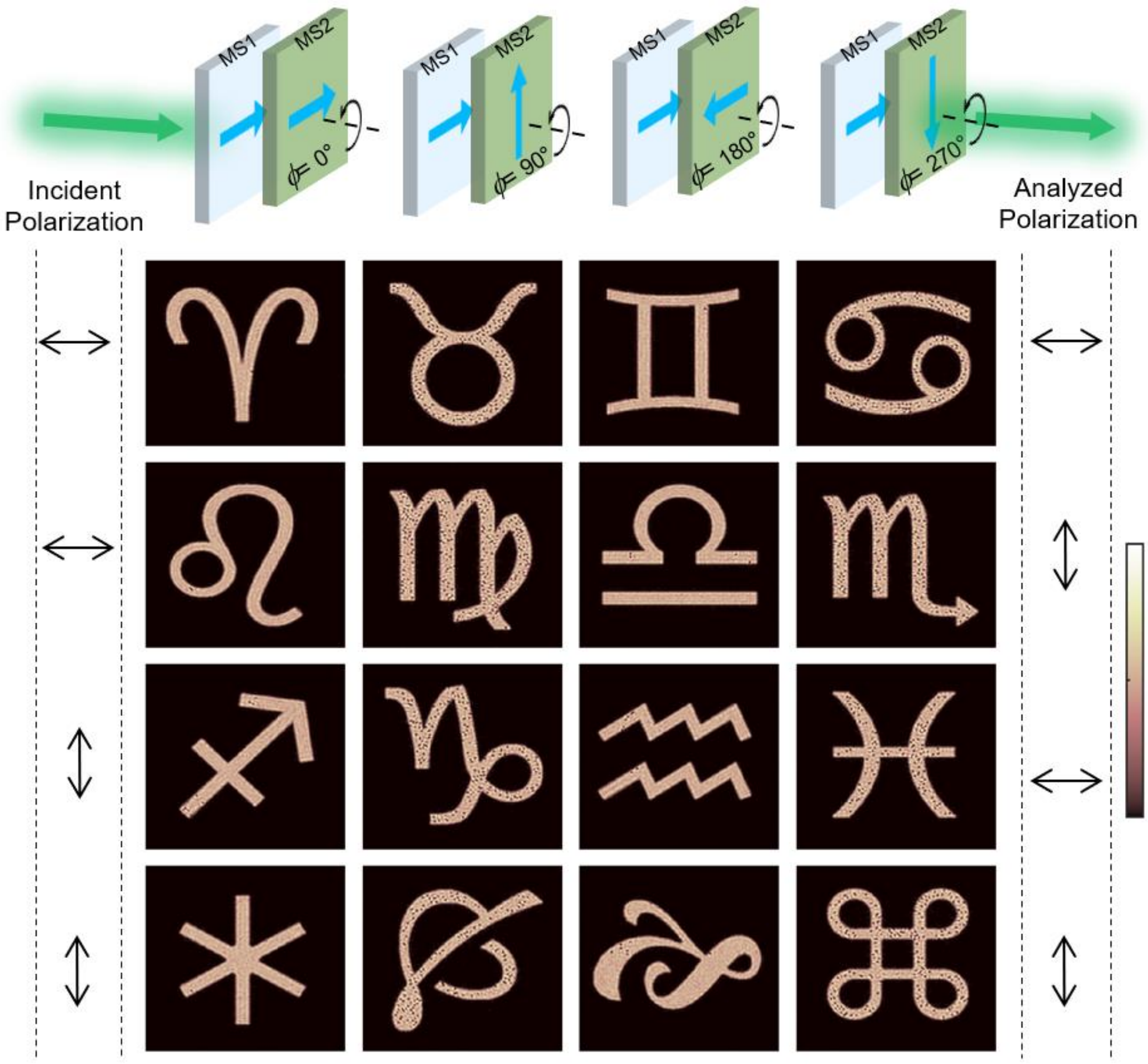

Figure S10. Optimized holographic images with the combinations of different rotation angles, incident polarizations and analyzed polarizations. The incident and analyzed polarizations are indicated at left side and right side, respectively.

## S11. Optical performance of the two layer metasurfaces designed with different gap distance

To show the performance of our design strategy with smaller gap distance, we choose several cases with smaller gap distance of $z$=5 μm, 10 μm , 20 μm and 100 μm, and the corresponding Jones matrixes of each single layer are calculated with gradient descent algorithm. The results are then simulated by FDTD, with the same procedures as that in supplementary section 6. Figure S11 shows the simulated nanoprintings and holographic images of the four cases. Basically, the qualities of holographic images are almost not influenced by the gap distance, while the fidelities of nanoprintings decrease as the gap distance reduces. However, even for the smallest distance of 5 μm, the simulated results show distinct nanoprintings as the designed ones, demonstrating the generality of our design strategy.

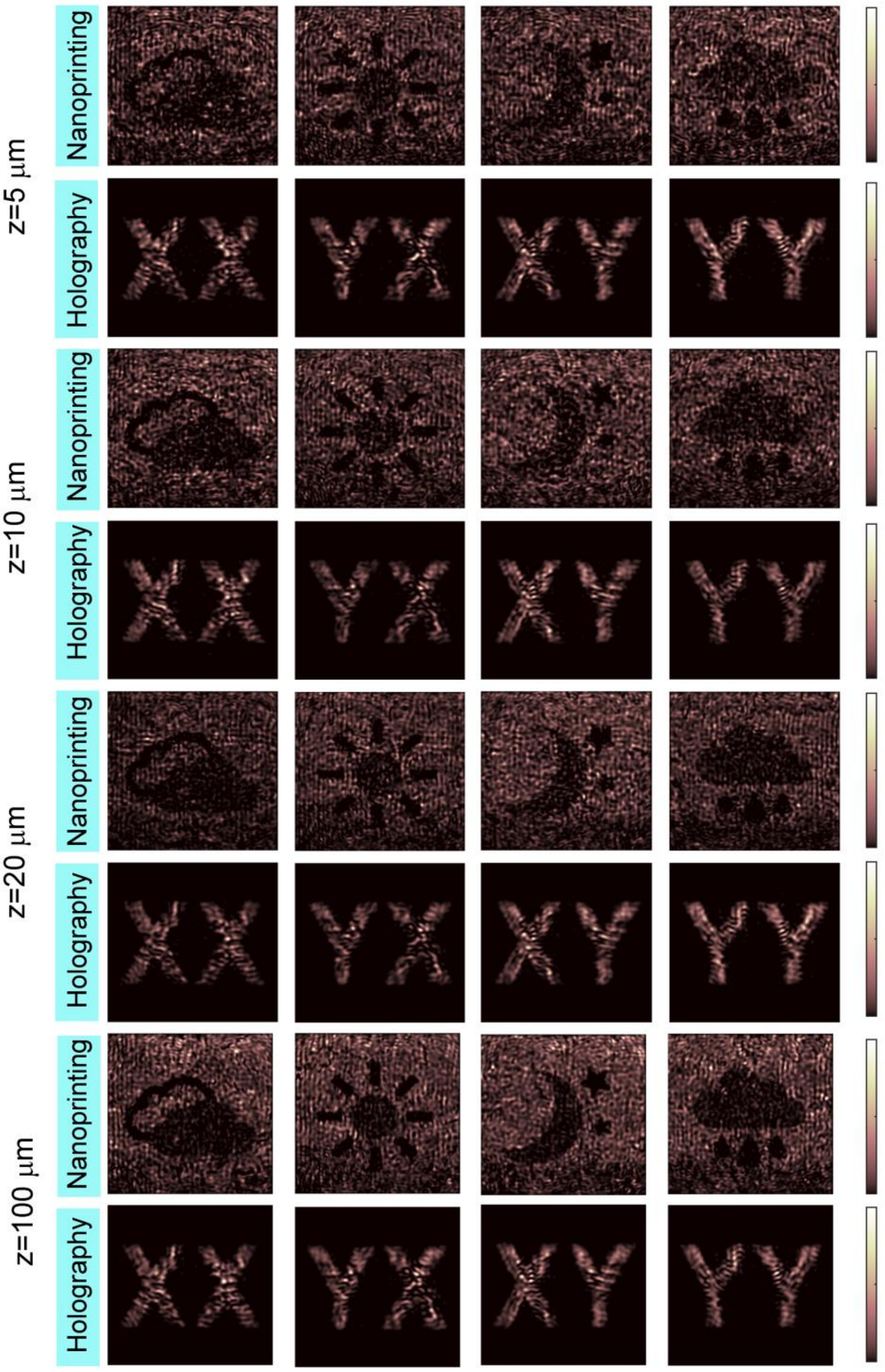

Figure S11. Simulated results of the nanoprinting and holographic images designed with different gap distances. The other geometric parameters are the same as that in supplementary section 6.